\documentclass{aa}

\usepackage{natbib}
\bibpunct{(}{)}{;}{a}{}{,} 

\usepackage{supertabular}
\usepackage{threeparttable}

\usepackage{soul} 
\usepackage{float}
\usepackage{graphicx}
\usepackage{caption}
\DeclareCaptionFormat{cont}{#1 (cont.)#2#3\par} 
\usepackage{txfonts}
\usepackage{siunitx}  
\usepackage{xcolor}
\usepackage{amsmath}
\usepackage{dblfloatfix}    

\usepackage[title]{appendix}

\newcommand{\tablenote}[1]{\parbox{18.3cm}{\indent \footnotesize{#1}}}
\newcommand{\tablenotea}[1]{\parbox{  8.9cm}{\indent \footnotesize{#1}}}

\begin{document} 

\title{Study of CS, SiO, and SiS abundances in carbon star envelopes: Assessing their role as gas-phase precursors of dust\thanks{Based on observations carried out with the IRAM 30 m Telescope. The Institut de Radioastronomie Millim\'etrique (IRAM) is supported by INSU/CNRS (France), MPG (Germany) and IGN (Spain).}}

\titlerunning{CS, SiO, and SiS abundances in carbon star envelopes}
\authorrunning{Massalkhi et al.}

\author{S.~Massalkhi, M.~Ag\'undez, and J.~Cernicharo}

\institute{Instituto de F\'isica Fundamental, CSIC, C/ Serrano 123, E-28006, Madrid, Spain}

\date{Received; accepted}


\abstract
{}
{We aim to determine the abundances of CS, SiO, and SiS in a large sample of carbon star envelopes covering a wide range of mass loss rates to investigate the potential role that these molecules could play in the formation of dust in the surroundings of the central AGB star.}
{We surveyed a sample of 25 carbon-rich AGB stars in the $\lambda$ 2 mm band, more concretely in the $J=3-2$ line of CS and SiO, and in the $J=7-6$ and $J=8-7$ lines of SiS, using the IRAM 30 m telescope. We performed excitation and radiative transfer calculations based on the large velocity gradient (LVG) method to model the observed lines of the molecules and to derive their fractional abundances in the observed envelopes. We also assessed the effect of infrared pumping in the excitation of the molecules.}
{We detected CS in all 25 targeted envelopes, SiO in 24 of them, and SiS in 17 sources. Remarkably, SiS is not detected in any envelope with a mass loss rate below $10^{-6}$ M$_\odot$ yr$^{-1}$ while it is detected in all envelopes with mass loss rates above that threshold. We found that CS and SiS have similar abundances in carbon star envelopes, while SiO is present with a lower abundance. We also found a strong correlation in which the denser the envelope, the less abundant are CS and SiO. The trend is however only tentatively seen for SiS in the range of high mass loss rates. Furthermore, we found a relation in which the integrated flux of the MgS dust feature at 30\,$\muup$m increases as the fractional abundance of CS decreases.}
{The decline in the fractional abundance of CS with increasing density could be due to gas-phase chemistry in the inner envelope or to adsorption onto dust grains. The latter possibility is favored by a correlation between the CS fractional abundance and the 30\,$\muup$m feature, which suggests that CS is efficiently incorporated onto MgS dust around C-rich AGB stars. In the case of SiO, the observed abundance depletion with increasing density is most likely caused by an efficient incorporation onto dust grains. We conclude that CS, SiO (very likely), and SiS (tentatively) are good candidates to act as gas-phase precursors of dust in C-rich AGB envelopes.}


\maketitle

\section{Introduction}

The circumstellar envelopes (CSEs) of asymptotic giant branch (AGB) stars, formed through extensive stellar mass loss, are rich in chemical diversity and have long been known to be efficient sites for molecule and grain formation. In the Galaxy, AGB stars are the main sources of interstellar dust \citep{geh1989}. It is therefore of particular interest to study their CSEs because these envelopes contain gas (mainly in the form of molecules), dust, and products of nucleosynthesis that are eventually expelled into the interstellar medium (ISM) and will contribute to its evolution and enrichment \citep{her2005}. 

The type of molecules and dust grains found in CSEs is to a great extent determined by the carbon-to-oxygen (C/O) abundance ratio at the photosphere of the AGB star. At the temperatures and densities of AGB atmospheres, thermochemical equilibrium (TE) predicts that carbon monoxide (CO) molecules have a very high abundance because of their stability, which locks most of the available carbon in oxygen-rich stars (M-type, C/O $<$ 1) or oxygen in carbon-rich stars (C-type, C/O $>$ 1). As a consequence of this, envelopes around M-type stars contain a variety of oxygen-bearing molecules (e.g., H$_2$O, SiO, and TiO; see \citealt{vel2017}) and silicate dust, while CSEs around C-type stars display a variety of carbon-bearing molecules (e.g., C$_2$H$_2$, HCN, CS, and SiC$_2$; see \citealt{olo1993} and \citealt{cer2000}) and contain carbonaceous, silicon carbide, and magnesium sulfide dust. After some gas-phase precursors condense near the surroundings of the stellar photosphere, the condensation nuclei grow to micrometer sizes as a consequence of accretion and coagulation processes. In C-rich AGB stars, molecules such as CS \citep{olo1993}, SiS \citep{sch2007}, and SiO \citep{sch2006a} are found to be abundant and are potential precursors of dust grains, in which case their abundances in the gas phase should experience a decline as they incorporate into solid grains. Eventually, interstellar ultraviolet (UV) photons destroy the molecules remaining in the gas phase in the outer envelope owing to photodissociation. However, the picture is yet poorly constrained from an observational point of view. It is still not well understood what are the gas-phase building blocks of dust grains in CSEs. 

\begin{table*}
\caption{Sample of carbon stars}\label{table:sources}
\centering
\resizebox{\linewidth}{!}{
\begin{tabular}{lrrcrcrccccc}
\hline \hline
\multicolumn{1}{l}{Name} & \multicolumn{1}{c}{R.A.} & \multicolumn{1}{c}{Dec.} & \multicolumn{1}{c}{$V_{\rm LSR}$} & \multicolumn{1}{c}{$D$} & \multicolumn{1}{c}{$T_{\star}$} & \multicolumn{1}{c}{$L_{\star}$} & \multicolumn{1}{c}{$\dot{M}$} & \multicolumn{1}{c}{$V_{\rm exp}$} &   \multicolumn{1}{c}{$T_{\rm d}(r_c)$} & \multicolumn{1}{c}{$r_c$} & $\Psi$ \\

\multicolumn{1}{c}{}          & \multicolumn{1}{c}{J2000.0} & \multicolumn{1}{c}{J2000.0} & \multicolumn{1}{c}{(km~s$^{-1}$)} & \multicolumn{1}{c}{(pc)} & \multicolumn{1}{c}{(K)} & \multicolumn{1}{c}{(L$_{\odot}$)}  & \multicolumn{1}{c}{(M$_{\odot}$ yr$^{-1}$)} & \multicolumn{1}{c}{(km~s$^{-1}$)} &\multicolumn{1}{c}{(K)} & \multicolumn {1}{c}{(cm)} &   \\
\hline
IRC\,+10216 & 09:47:57.45 & $+$13:16:43.9 & $-26.5$ & 130   & 2330   &  8750   & $2.0\times10^{-5}$  & 14.5 & 800   &  $2.0\times10^{14}$  &  300    \\
CIT\,6       & 10:16:02.27 & $+$30:34:18.6 & $-1$    & 400  & 1800   & 10000  & $6.0\times10^{-6}$  & 17    & 1000   & $2.1\times10^{14}$  & 141       \\
CRL\,3068    & 23:19:12.24 & $+$17:11:33.4 & $-31.5$ & 1300 & 1800   & 10900  & $2.5\times10^{-5}$  & 14.5  & 1500   & $2.0\times10^{14}$  & 174    \\ 
S\,Cep       & 21:35:12.83 & $+$78:37:28.2 & $-15.3$ & 380  & 2200   & 7300   & $1.2\times10^{-6}$  & 22.5  & 1400   & $5.8\times10^{13}$  & 360    \\
IRC\,+30374  & 19:34:09.87 & $+$28:04:06.3 & $-12.5$ & 1200 & 2000   & 9800   & $1.0\times10^{-5}$  & 25    & 1000   & $2.2\times10^{14}$  & 1008   \\
Y\,CVn       & 12:45:07.83 & $+$45:26:24.9 & $+22$   & 220  & 2200   & 4400   & $1.5\times10^{-7}$  & 7     & 1500   & $8.7\times10^{13}$  & 500     \\
LP\,And      & 23:34:27.53 & $+$43:33:01.2 & $-17$   & 630  & 1900   & 9600   & $7.0\times10^{-6}$  &14.5   & 1100   & $1.8\times10^{14}$  & 288     \\ 
V\,Cyg       & 20:41:18.27 & $+$48:08:28.8 & $+13.5$ & 366  & 2300   & 6000   & $1.6\times10^{-6}$  & 12    & 1400   & $9.4\times10^{13}$  & 364      \\
UU\,Aur      & 06:36:32.84 & $+$38:26:43.8 & $+6.7$  & 260  & 2800   & 6900   & $2.4\times10^{-7}$  & 10.6  & 1500   & $6.3\times10^{13}$  & 1000 \\ 
V384\,Per    & 03:26:29.51 & $+$47:31:48.6 & $-16.8$ & 560  & 2000   & 8100   & $2.3\times10^{-6}$  & 15.5  & 1300   & $1.0\times10^{14}$  & 584     \\
IRC\,+60144  & 04:35:17.54 & $+$62:16:23.8 & $-48.8$ & 1030 & 2000   & 7800   & $3.7\times10^{-6}$  & 19.5  & 1200   & $2.0\times10^{14}$  & 1014  \\
U\,Cam       & 03:41:48.17 & $+$62:38:54.4 & $+6$    & 430  & 2695   & 7000   & $2.0\times10^{-7}$  & 13    & 1500   & $4.4\times10^{13}$  & 833     \\  
V636\,Mon    & 06:25:01.43 & $-$09:07:15.9 & $+10$   & 880  & 2500   & 8472   & $5.8\times10^{-6}$  & 20    & 1200   & $1.7\times10^{14}$  & 300    \\
IRC\,+20370  & 18:41:54.39 & $+$17:41:08.5 & $-0.8$  & 600  & 2200   & 7900   & $3.0\times10^{-6}$  & 14    & 1500   & $8.1\times10^{13}$  & 266    \\
R\,Lep       & 04:59:36.35 & $-$14:48:22.5 & $+11.5$ & 432  & 2200   & 5500   & $8.7\times10^{-7}$  & 17.5  & 1000   & $1.8\times10^{14}$  & 500  \\ 
W\,Ori       & 05:05:23.72 & $+$01:10:39.5 & $-1$    & 220  & 2600   & 3500   & $7.0\times10^{-8}$  & 11    & 1500   & $4.3\times10^{13}$  & 333     \\
CRL\,67      & 00:27:41.10 & $+$69:38:51.5 & $-27.5$ & 1410 & 2500   & 9817   & $1.1\times10^{-5}$  & 16    & 1200   & $1.8\times10^{14}$  & 495   \\
CRL\,190     & 01:17:51.62 & $+$67:13:55.4 & $-39.5$ & 2790 & 2500   & 16750  & $6.4\times10^{-5}$  & 17    & 1000   & $4.7\times10^{14}$  & 424     \\
S\,Aur       & 05:27:07.45 & $+$34:08:58.6 & $-17$   & 300  & 3000   & 8900   & $4.0\times10^{-7}$  & 24.5  & 1500   & $7.3\times10^{13}$  & 500     \\ 
V\,Aql       & 19:04:24.15 & $-$05:41:05.4 & $+53.5$ & 330  & 2800   & 6500   & $1.4\times10^{-7}$  & 8     & 1500   & $6.1\times10^{13}$  & 500      \\
CRL\,2513    & 20:09:14.25 & $+$31:25:44.9 & $+17.5$ & 1760 & 2500   & 8300   & $2.0\times10^{-5}$  & 25.5  & 1200   & $1.6\times10^{14}$  & 453      \\
CRL\,2477    & 19:56:48.43 & $+$30:43:59.9 & $+5$    & 3380 & 3000   & 13200  & $1.1\times10^{-4}$  & 20    & 1800   & $2.8\times10^{14}$  & 532       \\
CRL\,2494    & 20:01:08.51 & $+$40:55:40.2 & $+29$   & 1480 & 2400   & 10200  & $7.5\times10^{-6}$  & 20    & 1200   & $2.3\times10^{14}$  & 436     \\
RV\,Aqr      & 21:05:51.74 & $-$00:12:42.0 & $+0.5$  & 670  & 2200   & 6800   & $2.3\times10^{-6}$  & 15    & 1300   & $7.6\times10^{13}$  & 200     \\
ST\,Cam      & 04:51:13.35 & $+$68:10:07.6 & $-13.6$ & 360  & 2800   & 4400   &  $1.3\times10^{-7}$ & 8.9   & 1500   & $5.0\times10^{13}$  & 500       \\ \hline
\end{tabular}                                                                                               
}
\tablenote{\\
The adopted parameters are discussed in \cite{mas2018}. See references therein. $\Psi$ for UU\,Aur and R\,Lep is from \cite{sch2001} and for IRC\,+60144 from \cite{gro2002}}
\end{table*}

One of the first major studies of abundances in a large sample of AGB stars was performed by \cite{gon2003}, who investigated SiO in $\sim$40 M-type stars. Later on, \cite{sch2006a} studied SiO in a sample of 19 C-rich AGB stars. Interestingly, it was found that SiO behaves similarly in both types of stars, showing a trend of decreasing abundance with increasing mass loss rate, thought to be due to an increased depletion of SiO onto dust grains. On the other hand, when \cite{sch2007} investigated SiS in a reduced sample of C-rich stars, they did not find a clear trend; this contrasted with the results of SiO, which could imply that SiS is less likely to be adsorbed onto dust grains than SiO in carbon-rich envelopes. Recently, we investigated SiC$_2$ in a sample of 25 carbon-rich AGB stars and found a similar trend as that found for SiO; that is, we discovered an abundance decline with increasing envelope density, which points to SiC$_2$ being efficiently incorporated into dust grains and playing an important role in the formation of silicon carbide dust \citep{mas2018}. 

In this paper, we follow up on our last study to investigate the abundance of CS, SiO, and SiS in the envelopes of carbon stars and to understand their potential role as gas-phase precursors of dust grains. We present observations of SiO ($J=3-2$), SiS ($J=7-6$ and $J=8-7$), and CS ($J=3-2$) in a sample of 25 carbon stars with diverse mass loss rates. We carried out a detailed non-local thermodynamic equilibrium (non-LTE) radiative transfer analysis to derive molecular abundances in the CSEs. The sample of stars and observational details are presented in Sec.~\ref{sec:observations} and the main results obtained from the observations in Sec.~\ref{sec:results}. In Sec.~\ref{sec:model} we describe the model and the excitation and radiative transfer calculations and discuss the most interesting features from these calculations in Sec.~\ref{sec:results_model}. Finally, we discuss the main implications of our study in Sec.~\ref{sec:discussion} and present our conclusions in Sec.~\ref{sec:conclusions}.

\section{Observations} \label{sec:observations}

The observations were carried out in September 2017 with the IRAM 30 m telescope, located at Pico Veleta, Spain. The sample of 25 C-rich AGB stars observed is the same used in our previous study of SiC$_2$ \citep{mas2018} and was selected according to intense molecular emission, mainly based on the intensity of the HCN $J=1-0$ line \citep{lou1993,buj1994,sch2013}. The observed sources and their parameters are listed in Table~\ref{table:sources}. In this study, we focused on the emission of CS, SiO, and SiS and therefore the spectral setup used was slightly shifted from that employed in \cite{mas2018} and accommodated in a way to include the lines CS $J=3-2$, SiO $J=3-2$ and SiS $J=7-6$ and $J=8-7$ (see line parameters in Table~\ref{table:lines}).

We used the E150 receiver in dual side band, with image rejections $>$10 dB, and observed the frequency ranges $125.7-133.5$ GHz and $141.4-149.2$ GHz in the lower and upper side bands, respectively. The beam size of the telescope at these frequencies is in the range 16.7-19.3$''$. We used the wobbler-switching technique. This technique consists of a symmetric position switching with the OFF position (atmosphere) at 180$''$ in azimuth from the ON position (source + atmosphere). Spectra at the OFF and ON positions are taken by nutating the secondary mirror at a rate of 0.5 Hz, and the OFF is subtracted from the ON to remove the contribution of the atmosphere to the signal. The focus was regularly checked on Venus and the pointing of the telescope was systematically checked on a nearby quasar before observing each AGB star. The error in the pointing is estimated to be 2-3$''$. The E150 receiver was connected to a fast Fourier transform spectrometer providing a spectral resolution of 0.2 MHz. The weather was good and stable during most of the observations, with typical amounts of precipitable water vapor of 1-3 mm and average system temperatures of 115 K. The intensity scale, calibrated using two absorbers at different temperatures and the atmospheric transmission model (ATM) \citep{cer1985,par2001}, is expressed in terms of $T_{\rm mb}$, the main beam brightness temperature. The error in the intensities due to calibration is estimated to be $\sim$20 \%. 

The data were reduced using CLASS\footnote{{\tiny Continuum and Line Analysis Single-dish Software}}\label{note:class} within the package GILDAS\footnote{\texttt{\tiny http://www.iram.fr/IRAMFR/GILDAS} \label{note:gildas}}. For each source, we averaged the spectra corresponding to the horizontal and vertical polarizations and subtracted a baseline consisting of a first order polynomial. When the lines were not very strong, the spectra were smoothed to a spectral resolution of 1 MHz to increase the signal-to-noise ratio. This corresponds to a velocity resolution of 2-2.4 km s$^{-1}$. Typical on source integration times, after averaging horizontal and vertical polarizations, were $\sim$1 h for each source, resulting in $T_{\rm mb}$ rms noise levels per 1 MHz channel of 2-6 mK.

\begin{table}
\caption{Covered rotational transitions of CS, SiO, and SiS} \label{table:lines}
\centering
\begin{tabular}{lccrr}
\hline \hline
\multicolumn{1}{l}{Transition}  & \multicolumn{1}{c}{Frequency} & \multicolumn{1}{c}{$A_{ul}$} & \multicolumn{1}{c}{$E_u$} & $\theta_{mb}$\\
\multicolumn{1}{l}{}                 & \multicolumn{1}{c}{(MHz)}  & \multicolumn{1}{c}{(s$^{-1}$)} & \multicolumn{1}{c}{(K)} & ($''$)\\
\hline
CS  $J=3-2$ & 146969.025 & $6.07\times10^{-5}$ & 14.1 & 16.7  \\
SiO $J=3-2$ & 130268.665 & $1.06\times10^{-4}$ & 12.5 & 18.8  \\
SiS $J=7-6$ & 127076.178 & $3.36\times10^{-5}$ & 24.4 & 19.3 \\
SiS $J=8-7$ & 145227.052 & $5.05\times10^{-5}$ & 31.4 & 16.9 \\
\hline
\end{tabular}
\end{table}

\section{Observational results} \label{sec:results}

\begin{figure*}
\centering
\includegraphics[width=\textwidth]{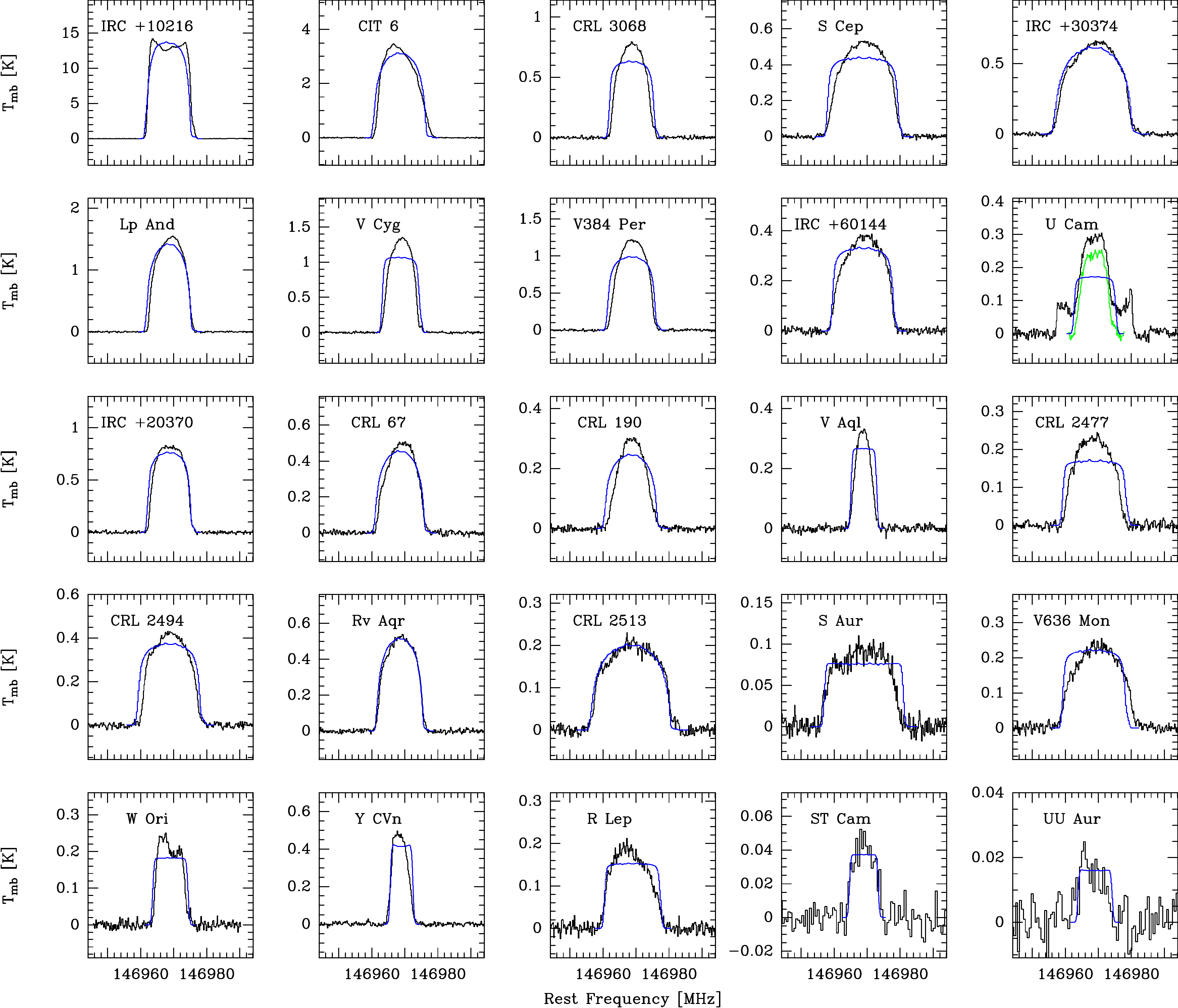}
\caption{CS $J=3-2$ line observed with the IRAM 30 m telescope in the 25 carbon stars (black histograms). U\,Cam shows emission contribution from the present day wind and the detached envelope. The green histogram corresponds to the observed line profile in which a fit to the detached envelope contribution (the wider one) has been subtracted. The blue lines indicate the calculated line profiles from the best-fit LVG model.}
\label{fig:cs_lines}
\end{figure*}
 
\begin{figure*}
\centering
\includegraphics[width=\textwidth]{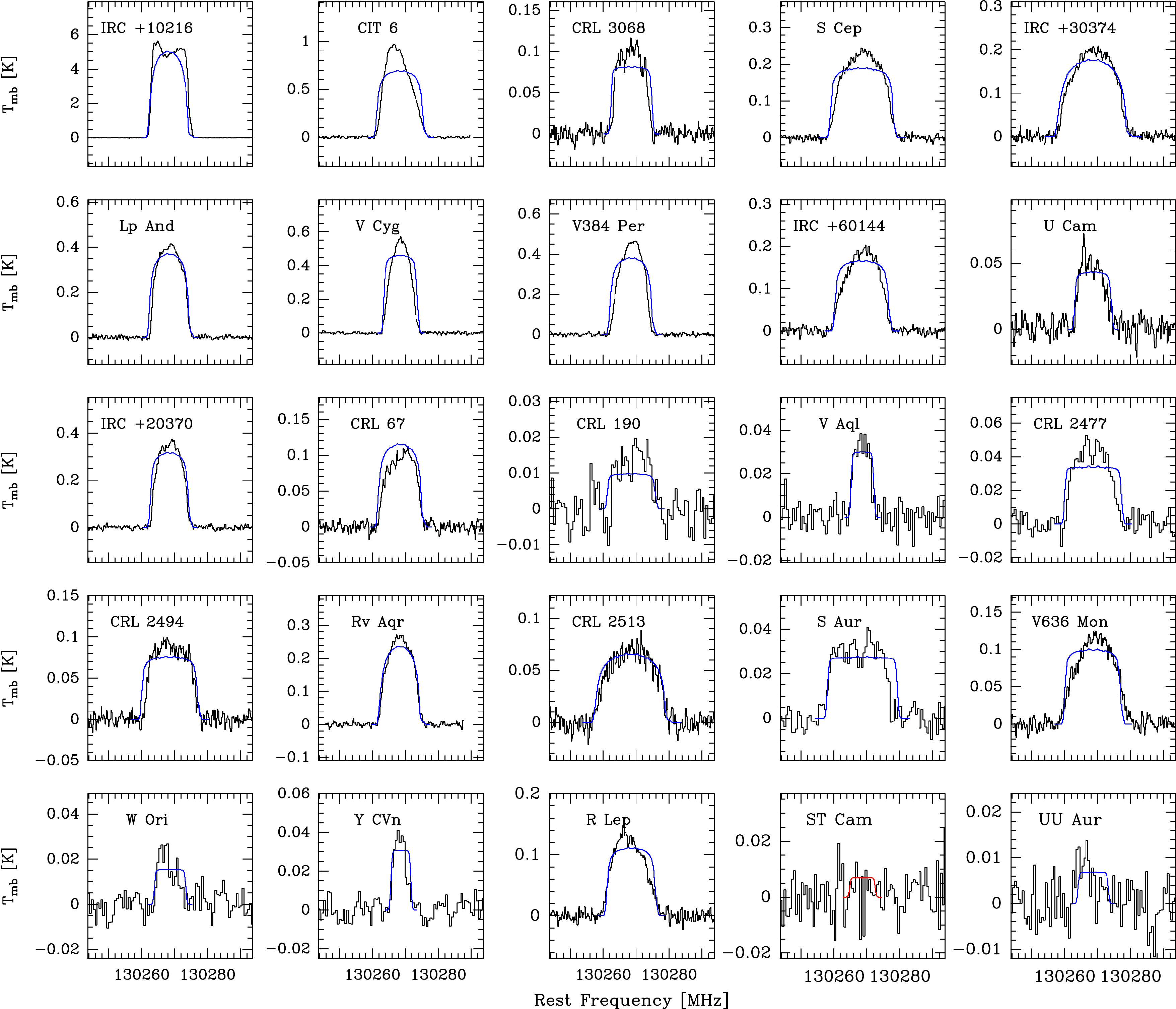}
\caption{SiO $J=3-2$ line observed with the IRAM 30 m telescope in the 25 carbon stars (black histograms). The blue lines indicate the calculated line profiles from the best-fit LVG model. The SiO line is not detected in ST\,Cam; the red line corresponds to the calculated line profile with the maximum intensity compatible with the non-detection.}
\label{fig:sio_lines}
\end{figure*}

\begin{figure*}
\centering
\includegraphics[width=\textwidth]{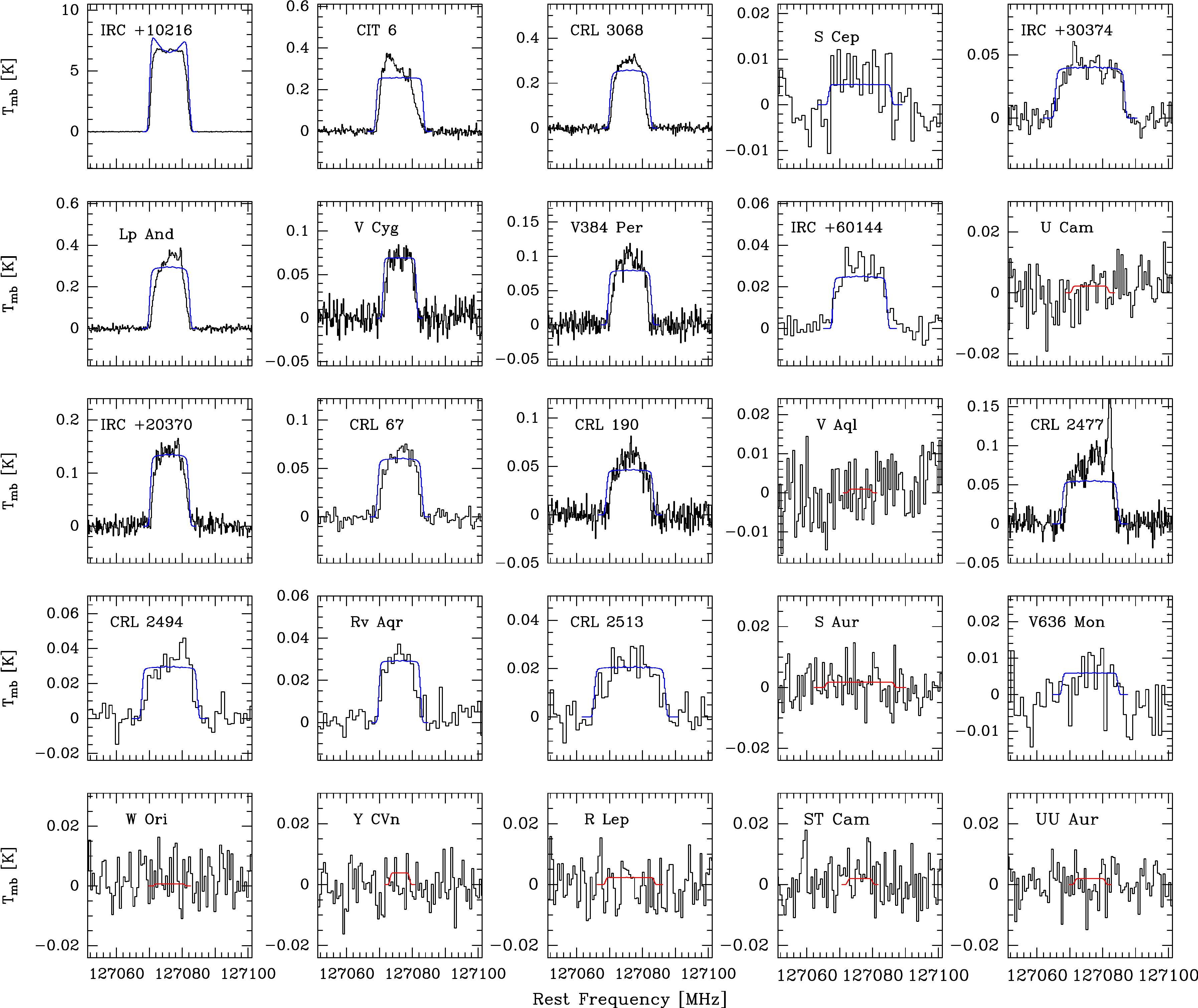}
\caption{SiS $J=7-6$ line observed with the IRAM 30 m telescope in the 25 carbon stars (black histograms). The blue lines indicate the calculated line profiles from the best-fit LVG model. The SiS line is not detected in various sources, for which we plot in red the calculated line profiles with the maximum intensity compatible with the non-detection.}
\label{fig:sis_7_6_lines}
\end{figure*}

\begin{figure*}
\centering
\includegraphics[width=\textwidth]{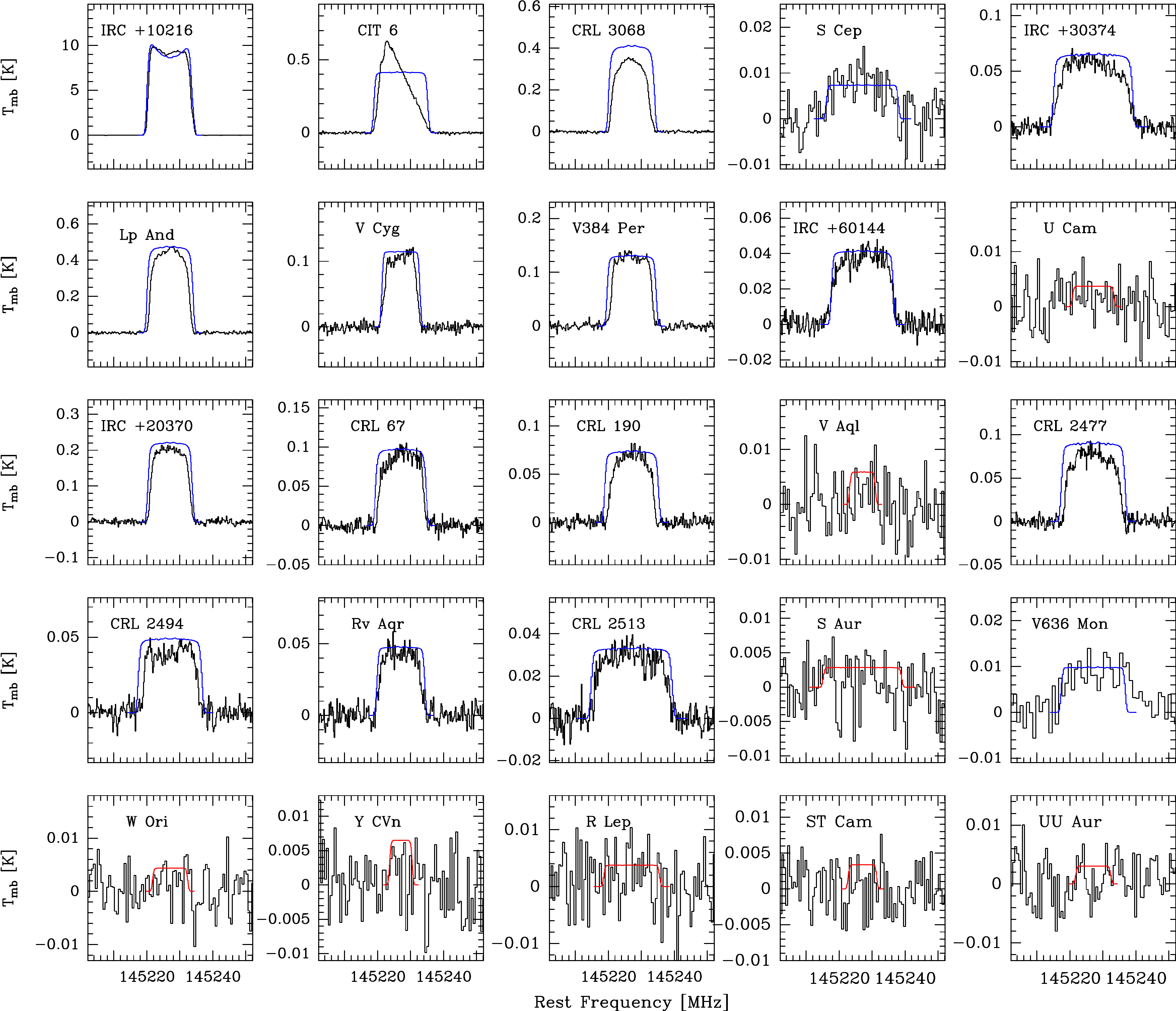}
\caption{Same as in Fig.~\ref{fig:sis_7_6_lines} but for SiS $J=8-7$.}
\label{fig:sis_8_7_lines}
\end{figure*}

\setcounter{table}{3}

The spectra obtained are shown in Fig.~\ref{fig:cs_lines} (CS $J=3-2$), Fig.~\ref{fig:sio_lines} (SiO $J=3-2$), and Figs.~\ref{fig:sis_7_6_lines} and \ref{fig:sis_8_7_lines} (SiS $J=7-6$ and $J=8-7$, respectively). The observed lines exhibit profiles typical of expanding circumstellar shells, i.e., either double-peaked, flat-topped, or parabolic, which can be reasonably well fit by the \texttt{shell} method of CLASS. The method fits the function

\begin{equation}\label{fig:shell_method}
f(\nu) = \frac{A}{\Delta \nu} \frac{1 + 4H[(\nu - \nu_{0})/\Delta \nu]^{2}}{1 + H/3}
,\end{equation}

where $A$ is the area under the profile in K MHz, $\nu_{0}$ is the middle frequency in MHz, $\Delta \nu$ is the full width at zero intensity level in MHz, and $H$ is the horn-to-center ratio, which is dimensionless. The expansion velocity V$_{\rm exp}$ can then be obtained by the following expression:

\begin{equation}
V_{\rm exp} = c \frac{\Delta \nu /2}{ \nu_{0}} 
,\end{equation}

where $c$ is the speed of light. \\ By performing the fit, we aim to derive for the target lines in every source the centroid frequency, expansion velocity, and line area, i.e., the velocity-integrated intensity. Most of the observed lines show profiles that match one of the types mentioned above, apart from the two sources U\,Cam (see line profile of CS emission in Fig.~\ref{fig:cs_lines}) and CIT\,6 (see line profile of SiO in Fig.~\ref{fig:sio_lines}), which have profiles that deviate from the expected line profile of a spherical expanding shell. The CS emission of U\,Cam is interpreted below as having two components, which we faced no difficulty considering separately. CIT\,6 shows a rather asymmetrical profile in the SiO $J=3-2$, SiS $J=7-6$ and $J=8-7$ lines. In any case, the two most interesting parameters, the line area and the expansion velocity, are reasonably well fitted. The derived parameters of the observed line profiles are shown in Table~\ref{table:line_parameters}.

The observations resulted in the clear detection of CS $J=3-2$ in all targeted sources. In the case of SiO, we detected the $J=3-2$ line in all target sources with the exception of ST\,Cam. As previously mentioned, the case of U\,Cam deserves particular attention because this source exhibits emission arising from a present-day wind and from a geometrically thin detached shell surrounding the central star \citep{sch2005}. The detached shell is probably the result of episodic mass loss driven by helium shell flashes, i.e., thermal pulses \citep{olo1990}. According to the observed line profiles, CS emission in U\,Cam arises from both the present-day wind and the detached envelope, while SiO emission arises exclusively from the present-day wind; compare line profiles of CS and SiO in Figs.~\ref{fig:cs_lines} and \ref{fig:sio_lines}, respectively, with that of CO $J=1-0$ in Fig. 6 of \cite{sch2005}. To disentangle the contribution from each component in the emission line of CS, we fitted the observed line with two components of different line widths; the narrow line width corresponds to the present-day wind and the wide to the detached envelope. The green line in Fig.~\ref{fig:cs_lines} corresponds to the emission from the present-day wind, which we are ultimately interested in modeling. In the case of SiS, the $J=7-6$ and $J=8-7$ lines were detected in 17 out of 25 sources, that is, in all sources except U\,Cam, V\,Aql, S\,Aur, W\,Ori, Y\,CVn, R\,Lep, ST\,Cam, and UU\,Aur.

The lines profile shapes usually give information about the emission being observed. When the emission is optically thick and unresolved by the beam of the telescope, the line profiles can be described as parabolic. We see that in most sources, the observed line shapes of CS and SiO exhibit such profile. If the emission is optically thin and unresolved, a flat-topped profile is seen. This kind of profile is seen in the observed lines of SiS. One notable exception for the three molecules is IRC\,+10216, whose close proximity and high mass loss rate result in an extended envelope whose molecular emission is spatially resolved by the telescope beam, and the line profiles show a more or less marked double-peak character.

\section{Excitation and radiative transfer modeling} \label{sec:model}

We aim at deriving molecular abundances of CS, SiO, and SiS in each source of our sample to provide a global view of how abundant these molecules are in envelopes around carbon stars. For this purpose, we performed excitation and radiative transfer calculations. The rotational lines of CS, SiO, and SiS studied in this work have upper level energies in the range 12.5 - 31.4 K. As shown in Sec.~\ref{sec:results_model}, the emission from these lines does not arise from the inner parts of the envelope but from the intermediate and outer regions, where gas densities are not high enough to thermalize the rotational levels. Therefore, level populations cannot be described by local thermodynamic equilibrium (LTE), and detailed non-LTE excitation and radiative transfer calculations have to be carried out taking into account the specific parameters for each envelope (see Table~\ref{table:sources}) to derive accurate molecular abundances.

We consider an envelope model consisting of a central AGB star surrounded by a spherically symmetric envelope of gas and dust expanding at constant velocity $V_{\rm exp}$ and with a mass loss rate $\dot{M}$, so that the radial density distribution follows an $r^{-2}$ law. The adopted physical properties of the stars and associated CSEs are presented in Table~\ref{table:sources}. The various physical quantities describing the envelope, such as the radial profiles of the gas density, gas temperature, and  dust temperature, as well as the properties of the dust grains, are described in \cite{mas2018}. 

We performed excitation and radiative transfer calculations to model the line emission of the studied molecules based on the multishell large velocity gradient (LVG) method. The LVG formalism is described in \cite{sob1960} and \cite{cas1970}, whereas the implementation of the multishell method for CSEs is described in more detail in \cite{agu2009} and \cite{agu2012}. The LVG method deals well with optically thin lines and lines with moderate optical thickness (Castor 1970). This method is a good compromise with respect to other methodologies such as Monte Carlo, which are more computationally expensive and exhibit problems of convergence when including a high number of energy levels. Briefly, the circumstellar envelope is divided into a number of concentric shells, each of which has a characteristic set of physical properties. The excitation and radiative transfer is solved locally in each shell through statistical equilibrium equations, where collisional and radiative processes determine the level populations. In each shell, the contribution of the background radiation field is included and is composed of the cosmic microwave background, stellar radiation, and thermal emission from dust. We also include infrared (IR) pumping, that is, absorption of IR photons and pumping to excited vibrational states followed by spontaneous radiative decay to rotational levels in the ground vibrational state. This process has been found to play an important role in the excitation of some molecules in IRC\,+10216 (e.g., \citealt{agu2006}). For the three molecules studied here (CS, SiO, and SiS), we only considered IR pumping to the first vibrationally excited state ($v=1$). We did verify that adding upper vibrational states had no effect on the calculated line intensities, i.e., including up to $v=1$ changed the line intensities for the three molecules in all the stars, but including up to $v=3$ did not cause further changes.

\subsection{Molecular data}

A major prerequisite for a successful radiative transfer code is the availability of accurate spectroscopic and collisional excitation data. We discuss below the spectroscopic and collisional excitation data of the three molecules that were input into our calculations. We considered enough rotational states to include levels with energies higher than 2000 K to better deal with the inner hot regions of the envelopes. 

In the excitation analysis of CS we considered the first 50 rotational levels within the $v=0$ and $v=1$ vibrational states (i.e., a total number of 100 energy levels). The level energies and transition frequencies were calculated from the Dunham coefficients given by \cite{mul2005}. The line strengths of pure rotational transitions were computed from the dipole moments for each vibrational state, $\mu_{v=0}$ =1.958 D and $\mu_{v=1}$ = 1.936 D \citep{win1968}, while for ro-vibrational transitions we used the Einstein coefficient of 15.8 s$^{-1}$ given for the $v=1\rightarrow0$ P(1) transition by \cite{cha1995}. We adopted the collision rate coefficients recently calculated by \cite{den2018} for pure rotational transitions and up to temperatures of 300 K. At higher temperatures and for ro-vibrational transitions we used the rate coefficients calculated by \cite{liq2007} multiplying the original values computed for He as collider by the squared ratio of the reduced masses of the H$_2$ and He colliding systems.

In the case of SiO, we considered the first 50 rotational levels of the ground and first excited vibrational states. To calculate the line frequencies and strengths, we used the Dunham coefficients given by \cite{sanz2003}, the dipole moments for pure rotational transitions within the $v=0$ and $v=1$ vibrational states of 3.0982 D and 3.1178 D, respectively, from \cite{ray1970} and an Einstein coefficient for the ro-vibrational transition $\nu=1\rightarrow0$ P(1) of 6.61 s$^{-1}$ from \cite{dri1997}. As collisional rate coefficients we adopted those calculated by \cite{day2006} for pure rotational transitions and for temperatures up to 300 K, while at higher temperatures and for ro-vibrational transitions we used the values computed by \cite{bal2017} scaling from He to H$_2$ as collider as in the case of CS.

For SiS, we include the first 70 rotational levels within the $v=0$ and $v=1$ vibrational states. Level energies were computed from the Dunham coefficients given by \cite{mul2007}. Line strengths were computed from the dipole moments $\mu_{v=0}$ =1.735 D, $\mu_{v=1}$ = 1.770 D, and $\mu_{v=1\rightarrow0}$ = 0.13 D \citep{mul2007,pin1987}. The collisional rate coefficients have been taken from the calculations of \cite{klos2008}, while for temperatures higher than 300 K and for ro-vibrational transitions we adopted the collisional rate coefficients computed by \cite{tob2008} scaled from He to H$_2$ as with CS and SiO.

\subsection{Abundance distributions} \label{subsec:emission_size}

We consider that CS, SiO, and SiS are formed close to the star with a given fractional abundance that remains constant throughout the envelope up to some region in the envelope where the abundance drops. This abundance falloff can be driven by at least two different processes: (1) condensation onto grains around the dust formation zone, and (2) photodissociation by the ambient UV radiation field in the outer envelope. While these molecules can certainly deplete in the dust formation region owing to condensation onto dust grains, in this work we are not sensitive to such potential abundance decline since the observed lines mostly probe intermediate and outer regions of the envelopes, that is, post-condensation regions (see Sec.~\ref{sec:results_model}). It is interesting to note that in IRC\,+10216, the emission from CS, SiO, and SiS vanishes at relatively outer radii, where photodissociation takes place (\citealt{bie1989,luc1992,luc1995}; Velilla-Prieto et al., submitted). Various studies have reported on the abundance depletion in the inner regions of IRC\,+10216. These studies show different degrees of depletion for the molecules, and in some cases the studies even have contradictory findings (\citealt{kea1993,boy1994,sch2006b,dec2010b,agu2012}.) Regardless of whether these molecules may experience a first abundance decline in the dust formation region or not, what is clear is that they maintain a significant abundance in the gas phase out to the outer envelope, where photodissociation is probably driving the disappearance of these molecules from the gas phase.

The above considerations suggest that a constant fractional abundance from the star and an abundance falloff driven by photodissociation in the outer envelope is a reasonable abundance distribution to model the lines observed in this work. In this scenario, the radial extent of each molecule would be entirely controlled by its corresponding photodissociation rate under the ambient UV radiation field and by the way in which circumstellar dust attenuates UV photons as they penetrate into the envelope. See \cite{mas2018} for more details on how the radial distributions were calculated using the photodissociation model. We however noticed that by using photodissociation rates from the literature when available (\citealt{hea2017} for SiO and \citealt{pat2018} for CS; see Sec.~\ref{sec:photodissociation} for more details) and adopting the canonical interstellar $N_{\rm H}$/$A_V$ ratio for the local ISM \citep{boh1978}, where $N_{\rm H}$ is the hydrogen column density in cm$^{-2}$ and $A_V$ is the visual extinction measured in magnitudes, the radial extent of these molecules is very likely underestimated, at least for some of the envelopes. This suspicion was based on the fact that the abundances derived for CS, SiO, and/or SiS were anomalously high in some sources, as they exceeded the elemental abundances of sulfur and/or silicon, which a priori are expected to be similar to those in the Sun \citep{asp2009}. Given the small number of observed lines (one for CS and SiO and two for SiS), the fact that they are sensitive to both the fractional abundance and the radial extent\footnote{{\tiny The emission arises from intermediate and outer regions of the envelopes and such regions are not resolved by the IRAM 30 m beam in most sources, with the notable exception of IRC\,+10216.}} and the fact that the radial extent is very likely not well described by a simple photodissociation model, we thus decided to fix the radial extent using an empirical correlation from the literature (see below) and leave as a free parameter the fractional abundance so that it can be derived by modeling the observed lines. 
Therefore, following the work by \cite{gon2003} and \citet{sch2006a,sch2007}, in this study we adopted a simple abundance distribution given by a Gaussian
\begin{equation}\label{eq:abundance_equation}
f(r) = f_0\,\exp{\Big(-(r/r_e)^{2}\Big)},
\end{equation}
where $f$ is the fractional abundance relative to H$_2$, $f_0$ is the initial abundance, and $r_e$ is the $e$-folding radius at which the abundance has dropped by a factor $e$. From a multiline study of SiO in M-type stars, \cite{gon2003} found the following empirical correlation between the $e$-folding radius and the envelope density evaluated through the quantity $\dot{M}/V_{\rm exp}$,
\begin{equation}\label{eq:scaling_law1}
\log r_{e}{\rm (SiO)} = 19.2 + 0.48 \log \left(\frac{\dot{M}}{V_{\rm exp}}\right), 
\end{equation}
where $r_{e}$ is given in cm, the mass loss rate $\dot{M}$ in M$_{\odot}$ yr$^{-1}$, and $V_{\rm exp}$ in km s$^{-1}$. Although the scaling law in Eq.~(\ref{eq:scaling_law1}) was derived for SiO in oxygen-rich stars, we adopted this law for SiO and SiS in our sample of carbon stars. \cite{sch2006a,sch2007} made the same assumption in the lack of better constraints for C-type stars. Concerning the assumption of similar radial extents for SiO and SiS, it is worth noting that although the photodissociation rate of SiS is unknown\footnote{\cite{pra1980} reported a photodissociation rate of $1.0\times10^{-10}$ s$^{-1}$ for SiS. Although it is difficult to trace the exact origin of this number, we suspect that it was most likely assumed to be equal to that of SiO.}, there are arguments that to a first order SiO and SiS should behave similarly with respect to photodissociation \citep{van1988,wir1994}. Interferometric observations of these two molecules in IRC\,+10216 show that they have similar emission sizes, where the SiO emission is slightly more extended than that of SiS (Velilla-Prieto et al., submitted). In the case of CS, we found it necessary to adopt a larger radial extent than for SiO and SiS because using Eq.~(\ref{eq:scaling_law1}) resulted in anomalously high CS abundances for some sources. We therefore adopted the following relation between $e$-folding radius and density in the envelope:
\begin{equation}\label{eq:scaling_law2}
\log r_{e}{\rm (CS)} = 19.65 + 0.48 \log \left(\frac{\dot{M}}{V_{\rm exp}}\right),
\end{equation}
which was derived by starting from Eq.~(\ref{eq:scaling_law1}) and increasing the first term in small steps so that the amount of sulfur locked in CS and SiS does not exceed the solar elemental abundance of sulfur (i.e., $f_0$(CS) + $f_0$(SiS) $\leq$ $2.6\times10^{-5}$; \citealt{asp2009}) in any envelope of our sample. Such a larger radial extent for CS, compared with SiO and SiS, is consistent with the lower photodissociation rate calculated for CS compared to that computed for SiO (\citealt{hea2017,pat2018}; see Sec.~\ref{sec:photodissociation}) and with the larger emission size observed for CS with respect to SiO and SiS in IRC\,+10216 (Velilla-Prieto et al., submitted).

\begin{figure}
\centering
\includegraphics[width=\columnwidth]{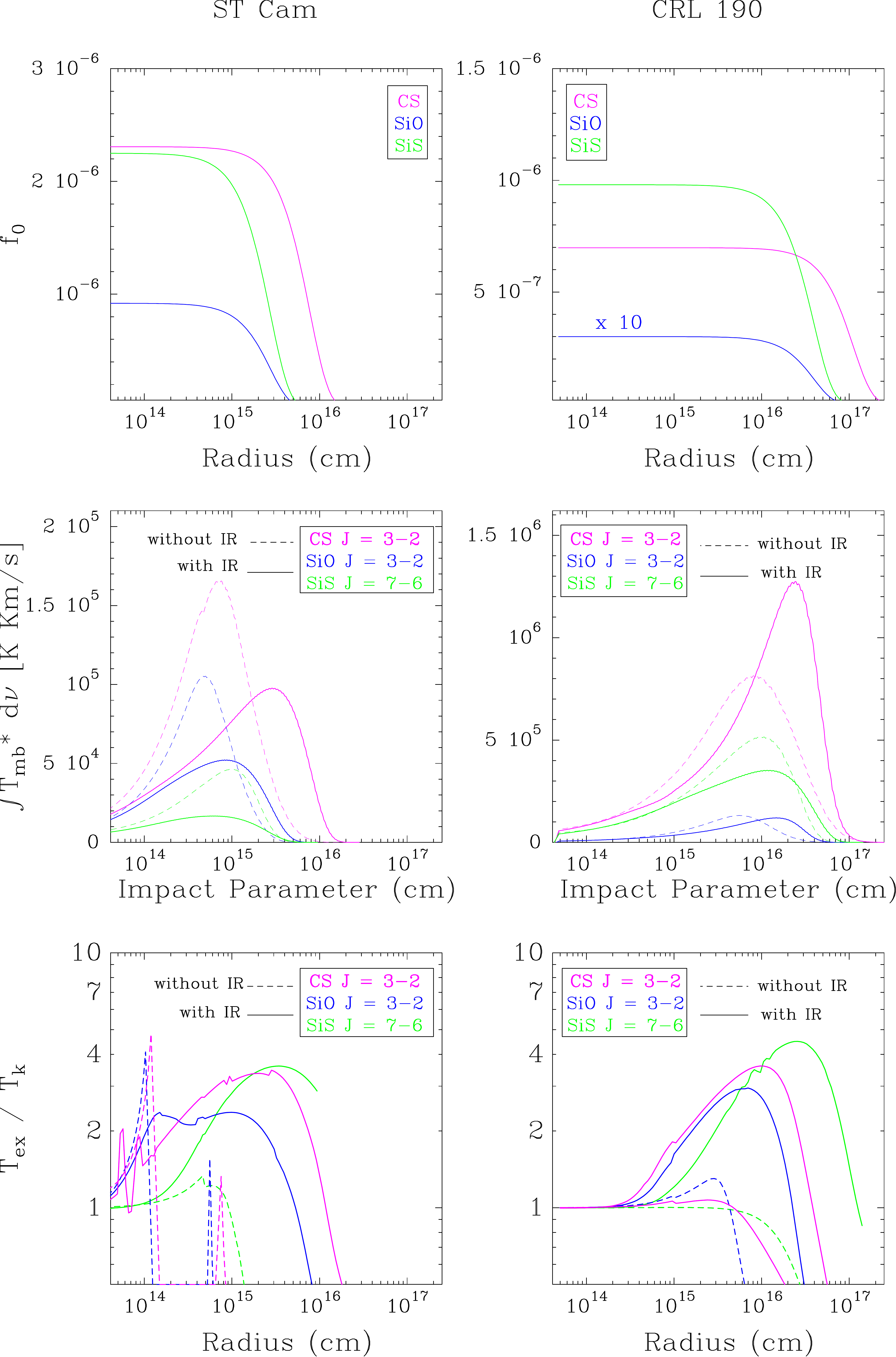}
\caption{Various features from the excitation and radiative transfer models for a low mass loss rate object (ST\,Cam) and a high mass loss rate object (CRL\,190). \textit{Top panels:} radial abundance profiles of CS, SiO, and SiS. \textit{Middle panels:} calculated velocity-integrated intensities for the observed lines of CS, SiO, and SiS as functions of the impact parameter. \textit{Bottom panels:} calculated ratio of excitation temperature to kinetic temperature ($T_{\rm ex}/T_{k}$) as a function of radius for observed lines of CS, SiO, and SiS. Solid lines correspond to the model in which IR pumping is included, whereas dashed lines correspond to the model excluding IR pumping. The SiS abundance in ST\,Cam corresponds to an upper limit.} \label{fig:ip}
\end{figure}

In a recent study, \cite{dan2018} derived empirical relations between $r_e$ and $\dot{M}$/$V_{\rm exp}$ for SiS and CS from a limited sample of M-, C-, and S-type stars. We noticed that implementing their empirical relations in our model calculations resulted in anomalously high abundances for CS and SiS for some low mass loss rate envelopes, for example, for W\,Ori $f_0$(CS) = $3.2\times10^{-5}$ and $f_0$(SiS) = $8.8\times10^{-5}$, which imply a sulfur abundance well above the solar value. The main lesson from these calculations is that the empirical relations derived by \cite{dan2018} cannot be safely extrapolated outside the relatively narrow range of $\dot{M}$/$V_{\rm exp}$ over which they were derived. We therefore decided not to adopt the CS and SiS scaling laws of these authors. It is worth noting that \cite{dan2018} derived larger $e$-folding radii for CS than for SiS, which again points to CS being more extended than SiS in agreement with the above arguments. Meanwhile, their SiS extent is smaller than that of SiO derived by \cite{gon2003}, which could point to SiS being photodissociated faster than SiO. \\

In summary, to model the emission lines of CS, SiO, and SiS and determine their abundance in the observed sources, we constructed a model of the envelope for each source with the parameters given in Table~\ref{table:sources} and adopting the abundance distribution given by Eq.~(\ref{eq:abundance_equation}), using Eq.~(\ref{eq:scaling_law1}) and Eq.~(\ref{eq:scaling_law2}) accordingly, and performed excitation and radiative transfer calculations by varying the initial fractional abundance relative to H$_2$, $f_0$ in Eq.~(\ref{eq:abundance_equation}), until the calculated line profiles matched the observed profiles. We chose the model that results in the best overall agreement between calculated and observed line profiles as the best-fit model. More specifically, our criterion was to match the line area of the calculated profile and the observed profile, and so we scaled the derived  abundance until the line area is reproduced. In those cases where no line is detected, we derive upper limits to the corresponding molecular abundance by choosing the maximum abundance that results in line intensities compatible with the noise level of the observations.

\section{Results from line modeling} \label{sec:results_model}

The calculated line profiles resulting from our best-fit LVG model for each of the sources are shown in blue in Fig.~\ref{fig:cs_lines} for CS, Fig.~\ref{fig:sio_lines} for SiO, and Figs.~\ref{fig:sis_7_6_lines} and~\ref{fig:sis_8_7_lines} for SiS, where they are compared with the observed line profiles (black histograms). We note that the overall agreement of the model is good given that our criterion of matching the line areas is fulfilled as described in the previous section. Concerning the line shapes, globally, the agreement is good as well.%
To facilitate the differentiation between line detections and non-detections, the calculated line profiles of the latter are plotted in red. In the case of CS in U\,Cam, the contribution from the present-day wind, which is that we are interested in modeling, is shown in green.

In Fig.~\ref{fig:ip} we show some salient features of the excitation and radiative transfer calculations for two envelopes, ST\,Cam and CRL\,190, which are representative of very different envelope densities. While ST\,Cam lies in the lower range, with $\dot{M}/V_{\rm exp}$ = $1.5\times10^{-8}$ M$_{\odot}$ yr$^{-1}$ km$^{-1}$ s, CRL\,190 lies at the higher end, with $\dot{M}/V_{\rm exp}$ = $3.7\times10^{-6}$ M$_{\odot}$ yr$^{-1}$ km$^{-1}$ s. The bottom panels of Fig.~\ref{fig:ip} show the calculated ratio of excitation temperature to gas kinetic temperature ($T_{\rm ex}/T_{k}$) for the observed lines of CS and SiO and that of the observed lines of SiS as a function of radius. We see that the rotational levels involved in the targeted transitions of the three molecules are thermalized ($T_{\rm ex}/T_{k}$ = 1) in the hot and dense inner regions, while as the radial distance from the star increases and the gas density decreases, the rotational levels deviate from thermalization. Concretely, lines become increasingly suprathermally excited ($T_{\rm ex}/T_{k}$ $>$ 1) with increasing radius, an effect that is largely caused by IR pumping. Therefore, IR pumping plays an important role in the excitation of the rotational transitions of these molecules. The much lower mass loss rate of ST\,Cam compared to CRL\,190 implies substantially lower densities in the envelope and thus in ST\,Cam rotational populations start to deviate from thermalization at shorter radii than in CRL\,190. The fact that the observed lines are not thermalized throughout most of the envelope, especially at low mass loss rates, stresses the need for non-LTE excitation calculations. For the low-mass loss rate ST\,Cam, the models without IR pumping for SiO and CS display a peculiar behavior caused by the excitation temperature becoming negative from intermediate regions of the envelope. Including IR pumping makes this behavior disappear. We note the SiS lines seem to end so abruptly because of the choice of the envelope outer edge in our model, where we chose the envelope to end at a radius at which the fractional abundance has dropped by a factor of 10$^{5}$.

The middle panels of Fig.~\ref{fig:ip} show the velocity-integrated intensity of the observed lines as a function of the impact parameter (solid lines). We see that in ST\,Cam the maximum contribution to the line emission comes from regions at a few $10^{15}$ cm from the star, while in the case of CRL\,190 the regions located at radii of a few $10^{16}$ cm contribute the most to the observed emission. Therefore, most of the emission of the observed lines arises from intermediate and outer regions of the envelope, rather than from inner regions. Moreover, the $\lambda$ 2 mm lines studied in this work probe regions where the abundance falloff has already started, which is especially marked in objects with low mass loss rates such as ST\,Cam (compare middle and top panels in Fig.~\ref{fig:ip}). This fact explains why the observed lines are sensitive to both the fractional abundance and the radial extent, and why adopting a more compact (extended) distribution would require a higher (lower) fractional abundance to reproduce the observed lines.

Infrared pumping not only plays an important role in regulating the excitation of the observed lines but also in determining their emission distribution. The bottom panels in Fig.~\ref{fig:ip} show that the emission is more compact if IR pumping is not included (dashed lines) than if it is taken into account (solid lines). Therefore, IR pumping results in a more extended emission distribution with an impact on the emerging line intensity. In fact, neglecting IR pumping results in a systematic decrease in the integrated line intensities of CS $J=3-2$ and SiO $J=3-2$ of $\sim60\%$ and $\sim50\%$, respectively, on average. In the case of SiS $J=7-6$ and $J=8-7$ the effect of excluding IR pumping is not as systematic as with CS and SiO as it leads to a decrease of the line intensities in some sources and an increase in other sources. If higher $J$ lines of SiS were targeted (above $J=10-9$) the effect of excluding IR pumping would be a systematic decrease in the integrated line intensities.

\section{Discussion} \label{sec:discussion}

\subsection{Fractional abundances}

\begin{table}
\caption{Fractional abundances of CS, SiO, and SiS derived}\label{table:abundances}
\centering
\resizebox{\columnwidth}{!}{
\begin{tabular}{lrcrrr}
\hline \hline
\multicolumn{1}{l}{Name}  & \multicolumn{1}{c}{$\dot{M}$} & \multicolumn{1}{c}{$V_{\rm exp}$}  & \multicolumn{1}{c}{$f_0$(CS)} & \multicolumn{1}{c}{$f_0$(SiO)} & \multicolumn{1}{c}{$f_0$(SiS)}\\
\multicolumn{1}{c}{}  & \multicolumn{1}{c}{(M$_{\odot}$ yr$^{-1}$)} & \multicolumn{1}{c}{(km~s$^{-1}$)} & & & \\
\hline
IRC\,+10216 & $2.0\times10^{-5}$   & 14.5  & $1.1\times10^{-6}$    & $2.6\times10^{-7}$   &    $1.3\times10^{-6}$      \\
CIT\,6      & $6.0\times10^{-6}$   & 17    & $6.4\times10^{-6}$    & $1.5\times10^{-6}$   &    $4.8\times10^{-6}$     \\
CRL\,3068   & $2.5\times10^{-5}$   & 14.5  & $7.4\times10^{-7}$    & $1.4\times10^{-7}$   &    $3.8\times10^{-6}$      \\ 
S\,Cep      & $1.2\times10^{-6}$   & 22.5  & $9.9\times10^{-6}$    & $9.4\times10^{-6}$   &    $1.9\times10^{-6}$      \\
IRC\,+30374 & $1.0\times10^{-5}$   & 25    & $1.0\times10^{-5}$    & $5.0\times10^{-6}$   &    $7.2\times10^{-6}$       \\
Y\,CVn      & $1.5\times10^{-7}$   & 7     & $5.7\times10^{-6}$    & $8.1\times10^{-7}$   &    <$8.0\times10^{-7}$   \\
LP\,And     & $7.0\times10^{-6}$   & 14.5  & $3.6\times10^{-6}$    & $1.0\times10^{-6}$   &    $7.0\times10^{-6}$       \\ 
V\,Cyg      & $1.6\times10^{-6}$   & 12    & $3.3\times10^{-6}$    & $2.9\times10^{-6}$   &    $4.0\times10^{-6}$       \\
UU\,Aur     & $2.4\times10^{-7}$   & 10.6  & $3.0\times10^{-7}$    & $2.5\times10^{-7}$  &    <$6.1\times10^{-7}$     \\ 
V384\,Per   & $2.3\times10^{-6}$   & 15.5  & $8.4\times10^{-6}$    & $6.4\times10^{-6}$   &    $1.0\times10^{-5}$       \\
IRC\,+60144 & $3.7\times10^{-6}$   & 19.5  & $7.3\times10^{-6}$    & $9.5\times10^{-6}$   &    $1.0\times10^{-5}$      \\
U\,Cam$^a$  & $2.0\times10^{-7}$   & 13    & $1.9\times10^{-5}$    & $1.0\times10^{-5}$   &    <$4.1\times10^{-6}$    \\  
V636\,Mon   & $5.8\times10^{-6}$   & 20    & $2.0\times10^{-6}$    & $1.7\times10^{-6}$   &    $9.6\times10^{-7}$      \\
IRC\,+20370 & $3.0\times10^{-6}$   & 14    & $4.1\times10^{-6}$    & $3.0\times10^{-6}$   &    $1.1\times10^{-5}$       \\
R\,Lep      & $8.7\times10^{-7}$   & 17.5  & $3.6\times10^{-6}$     & $5.7\times10^{-6}$    &    <$1.1\times10^{-6}$      \\  
W\,Ori      & $7.0\times10^{-8}$   & 11    & $2.1\times10^{-5}$    & $3.6\times10^{-6}$   &    <$4.9\times10^{-6}$    \\
CRL\,67     & $1.1\times10^{-5}$   & 16    & $3.5\times10^{-6}$    & $1.1\times10^{-6}$   &    $4.6\times10^{-6}$       \\
CRL\,190    & $6.4\times10^{-5}$   & 17    & $7.0\times10^{-7}$    & $3.0\times10^{-8}$   &    $9.8\times10^{-7}$       \\
S\,Aur      & $4.0\times10^{-7}$   & 24.5  & $9.3\times10^{-6}$    & $6.9\times10^{-6}$   &    <$3.3\times10^{-6}$     \\ 
V\,Aql      & $1.4\times10^{-7}$   & 8     & $1.0\times10^{-5}$    & $2.5\times10^{-6}$   &    <$2.3\times10^{-6}$     \\
CRL\,2513   & $2.0\times10^{-5}$   & 25.5  & $3.1\times10^{-6}$    & $1.3\times10^{-6}$   &     $2.8\times10^{-6}$     \\
CRL\,2477   & $1.1\times10^{-4}$   & 20    & $2.7\times10^{-7}$    & $1.0\times10^{-7}$   &     $1.7\times10^{-6}$     \\
CRL\,2494   & $7.5\times10^{-6}$   & 20    & $7.0\times10^{-6}$    & $2.9\times10^{-6}$   &     $1.1\times10^{-5}$     \\
RV\,Aqr     & $2.3\times10^{-6}$   & 15    & $7.7\times10^{-6}$    & $5.5\times10^{-6}$   &     $4.6\times10^{-6}$      \\
ST\,Cam     & $1.3\times10^{-7}$   & 8.9   & $2.3\times10^{-6}$    & <$9.2\times10^{-7}$  &     <$2.2\times10^{-6}$    \\ 
\hline
\end{tabular}}
\tablenotea{\small $^a$ U\,Cam has a present-day wind and a detached shell expanding away from the central star (for further details see \citealt{sch2005}). The values of $f_0$ given in this table correspond to the present-day wind.}
\end{table}

The abundances derived for CS, SiO, and SiS in the 25 carbon-rich envelopes studied are summarized in Table~\ref{table:abundances}. For some of these sources, abundances have been previously reported. For example, in IRC\,+10216 the abundances of CS, SiO, and SiS were derived by \cite{agu2012} from a multiline study including lines from vibrationally excited states. These authors found abundances relative to H$_2$ of $7\times10^{-7}$ for CS, $1.8\times10^{-7}$ for SiO, and $1.3\times10^{-6}$ for SiS, which are in very good agreement with the abundances derived in this work.

\cite{olo1993} looked for CS $J=2-1$ emission in a large sample of carbon stars, 12 of which are in our sample. However, in about half of these stars the line was not detected and only loose abundance upper limits could be derived. These authors used a relatively simple method for estimating the molecular abundances, which is based on an analytical expression and where they assumed a constant excitation temperature for simplicity. In those sources where CS was detected, the abundances derived were very high, often greater than the maximum amount obtainable for solar abundance of sulfur. They suggested that these anomalously high abundances probably originate from uncertainties in the envelope model. In another study, \cite{buj1994} surveyed a large sample of evolved stars in lines of several molecules including CS, SiO, and SiS. Their sample contains 16 C-rich AGB stars, 12 of which are also in our sample. These authors derived the abundances using an analytical expression that estimated only a lower limit if the line was optically thick. In general, their CS, SiO, and SiS abundances are lower than ours by a factor of a few and not higher than a factor of 10, apart from U\,Cam, where our derived CS and SiO abundances are one and two orders of magnitude, respectively, higher than theirs. \cite{buj1994} remarked that their abundances may be underestimated owing to optical depth effects.

More recently, \cite{sch2006a} surveyed a sample of 19 C-rich AGB stars and detected SiO line emission in all of these objects. They performed radiative transfer calculations to derive abundances adopting, similar to us, an abundance distribution based on the scaling law established by \cite{gon2003} for SiO in M-type stars. We share 13 sources with the former. Our SiO abundances are in good agreement with theirs for some sources, while for others there are significant differences. In LP\,And, for example, our derived SiO abundance is higher than theirs by almost a factor of ten, but on the other hand, our derived value in UU\,Aur is one order of magnitude lower than theirs, probably owing to differences in both the observations and the model. Later on, \cite{sch2007} reported on the detection of SiS line emission in 14 carbon stars, most of which are included in our sample. In general, the SiS abundances derived by us are similar to theirs, except for LP\,And, where we derive a SiS abundance higher than theirs by a factor of $\sim$6.

\begin{figure*}
\centering
\includegraphics[width=0.33\textwidth]{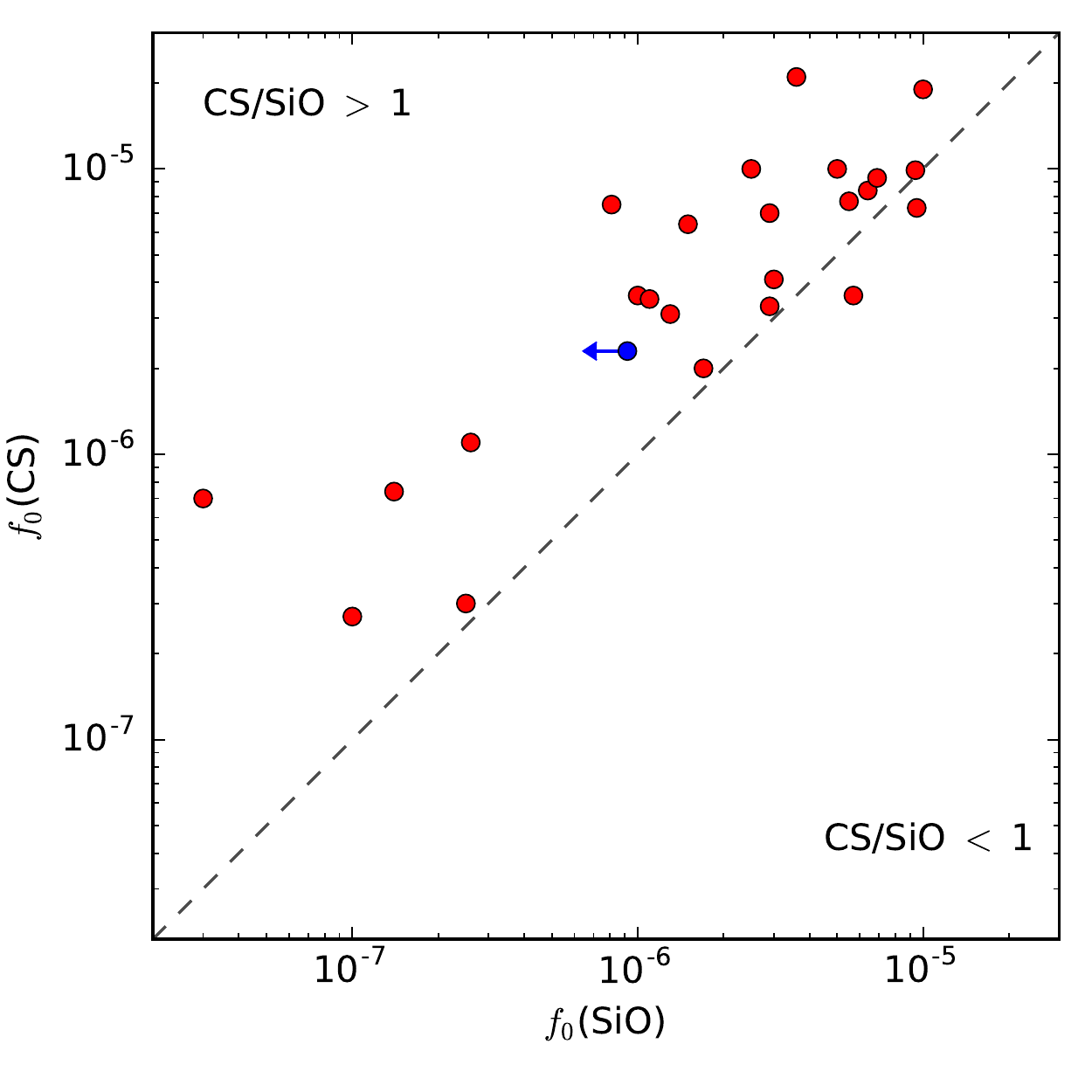} \includegraphics[width=0.33\textwidth]{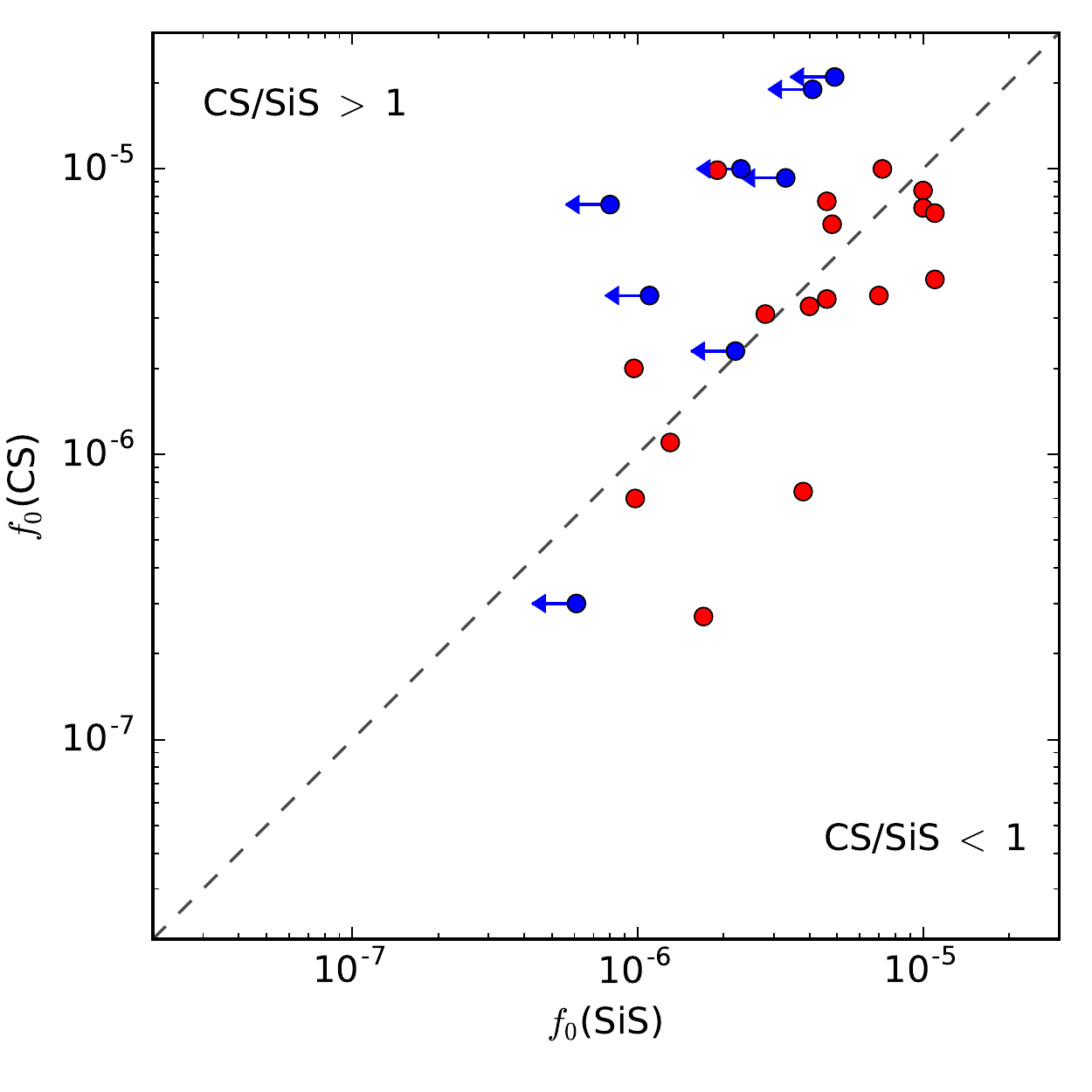} \includegraphics[width=0.33\textwidth]{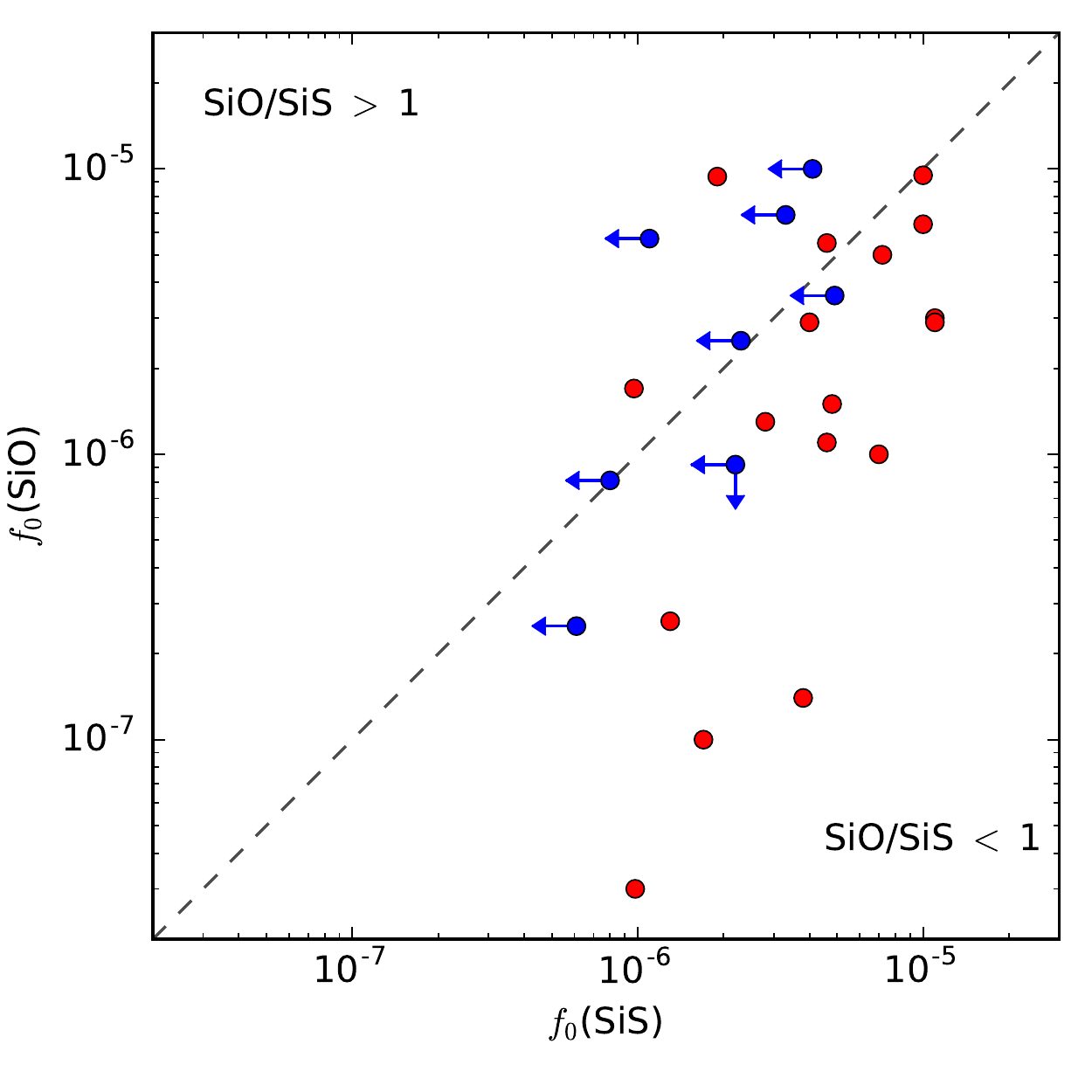}
\caption{Comparison of abundances between different pairs of molecules. The plots show the derived fractional abundances relative to H$_2$ of CS vs. SiO (left panel), CS vs. SiS (middle panel), and SiO vs. SiS (right panel). Those sources with non-detections are denoted with blue arrows. Diagonal dashed lines indicate where the abundances of the two molecules become equal.} \label{fig:ratios}
\end{figure*}

\begin{figure*}[b]
\centering
\includegraphics[width=\textwidth]{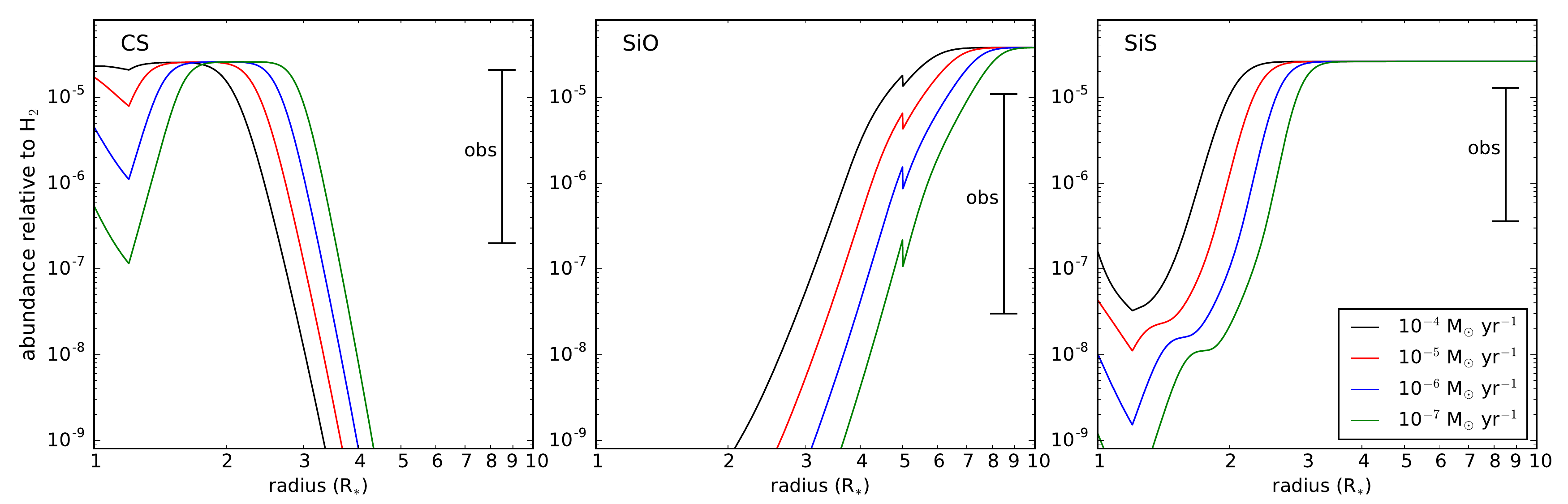}
\caption{Calculated fractional abundances of CS (left panel), SiO (middle panel), and SiS (right panel) under TE as a function of distance to the AGB star for various mass loss rates. The range of observed abundances is indicated.}\label{fig:lte_abun}
\end{figure*}

In this work, the fractional abundances relative to H$_2$ derived range between $2.7\times10^{-7}$ and $2.1\times10^{-5}$ for CS, $3\times10^{-8}$ and $1\times10^{-5}$ for SiO, and $<6.1\times10^{-7}$ and $1.1\times10^{-5}$ for SiS. Silicon monoxide is the molecule experiencing the largest variations (more than two decades) from source to source, while CS abundances span over two decades, and SiS shows the most uniform abundances across sources (with variations of less than two decades). We find that CS is systematically more abundant than SiO in most of the sources (see left panel of Fig.~\ref{fig:ratios}, where it is seen that most sources fall in the region of CS/SiO $>$ 1). We also note that SiO tends to be less abundant than SiS; the right panel of Fig.~\ref{fig:ratios} shows that for most sources SiO/SiS $\lesssim$ 1, while only in a few SiO/SiS is greater than one. Therefore, SiS seems to be a more efficient gas-phase reservoir of silicon than SiO in most carbon star envelopes. When comparing between CS and SiS, the two major gas-phase molecular reservoirs of sulfur in C-rich envelopes (see, \citealt{dan2018} and, e.g., the case of IRC\,+10216; \citealt{agu2012}), we notice that in some sources such as IRC\,+30374 and U\,Cam, CS is more abundant than SiS, while in others (e.g., CRL\,3068 and CRL\,2494) the contrary is found. Overall, the data points in the middle panel of Fig.~\ref{fig:ratios} fall along the line defined by $f_0$(CS) = $f_0$(SiS), with no clear preference for either the CS/SiS $>$ 1 or the CS/SiS $<$ 1 sides.

\begin{figure*}
\centering
\includegraphics[width=\textwidth]{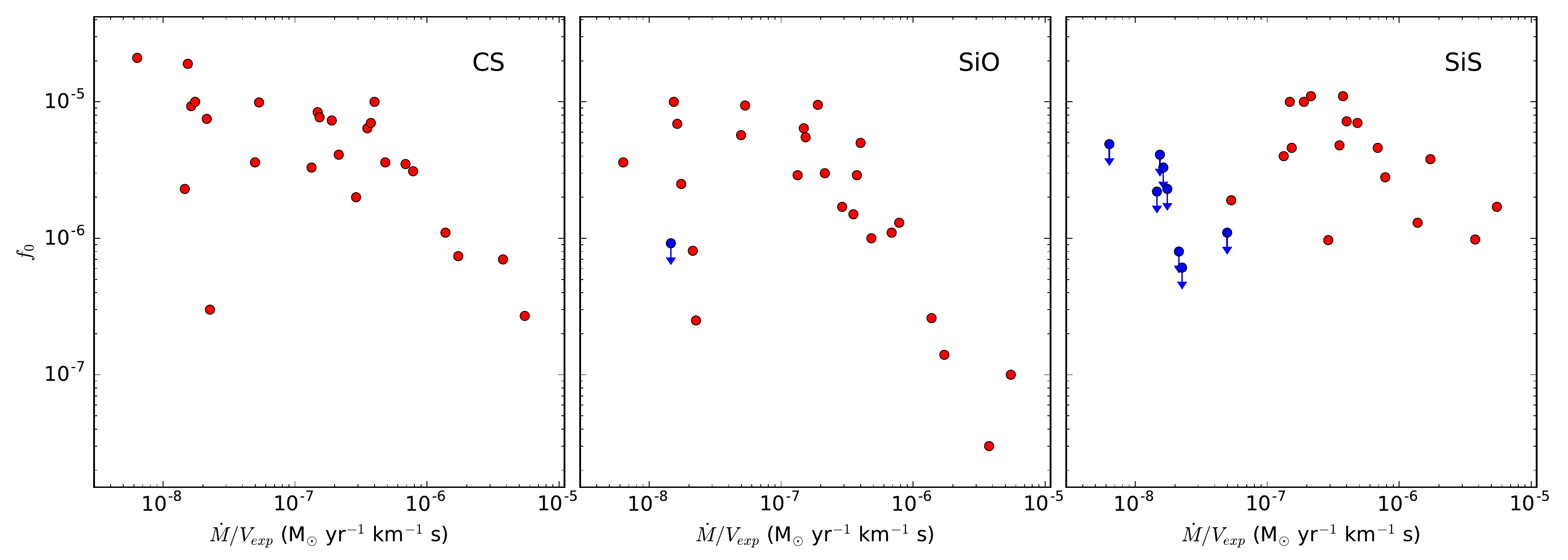}
\caption{Fractional abundances $f_0$ derived for CS (left panel), SiO (middle panel), and SiS (right panel) are plotted as a function of the envelope density proxy $\dot{M}/V_{\rm exp}$ for the 25 C-rich envelopes studied here. Blue downward arrows represent upper limits to $f_0$.} \label{fig:abun}
\end{figure*}

We can therefore extract as statistically meaningful conclusions that in carbon star envelopes CS and SiS have abundances of the same order, and that SiO is in general less abundant than CS and SiS. This is most likely a chemical feature related to C-rich AGB envelopes. \cite{dan2018} determined the CS and SiS abundances in a sample of AGB stars, and likewise found comparable CS and SiS abundances in C-rich AGB stars, while in their O-rich sample, SiS was found to be systematically higher than CS. S stars were also found to have similar abundances. In this line, it is interesting to see what  TE predicts for these molecules in C-rich AGB atmospheres. In Fig.~\ref{fig:lte_abun} we show the calculated fractional abundances of CS, SiO, and SiS as a function of radial distance from the star for various mass  loss rates. The underlying physical scenario adopted for the TE calculations is the same as adopted in \cite{mas2018}, where we used the radial profiles of density and temperature of IRC\,+10216 (\citealt{agu2012}; see downward revision on the density profile by \citealt{cer2013}) and scaled the density profile up or down depending on the mass loss rate, that is in the range of 10$^{-7}$-10$^{-4}$ M$_{\odot}$ yr$^{-1}$. Therefore, the TE calculations adopt different density profiles to account for low and high mass loss rates. We see that CS is predicted to have a high abundance (locking most of the available sulfur) in the close surroundings of the AGB star (at 1-3 $R_*$), while at farther distances its TE abundance vanishes to very low levels. On the other hand, SiS has almost the opposite behavior, with low abundance levels close to the star and a very high abundance beyond 2-3 $R_*$. The fact that CS and SiS are observed with similar abundances in our sample of envelopes is consistent with the abundances of these molecules being quenched to their TE values at radii of 2-3 $R_*$, which agrees with findings from the study of IRC\,+10216 \citep{agu2006,agu2012}. The process of abundance quenching is expected because as the gas expands the temperature and density drop and chemical reactions become slower and eventually too slow to proceed efficiently so that abundances are not further modified. Silicon monoxide shows a calculated radial TE abundance profile similar to that of SiS but shifted to larger radii. That is, SiO has a vanishingly small TE abundance close to the star, although calculations predict that this will reach a very high abundance (trapping most of the available silicon) beyond 5-8 $R_*$. The fact that the TE abundance of SiO is only high at a relatively large distance from the star is probably at the origin of the lower observed abundance of SiO compared to CS and SiS. In any case, given the range of SiO abundances derived from observations, SiO must quench its abundance to the TE value at somewhat larger radii than for CS and SiS. Predictions from a chemical kinetics model of the inner wind of AGB stars, including shocks driven by the pulsation of the star \citep{che2006}, indicate that for a C-rich object with a C/O elemental abundance ratio of 1.1, the abundances of CS, SiO, and SiS injected into the expanding envelope would be similar within a factor of $\sim1.5$, where CS is slightly more abundant than SiO and SiS. Given the scatter in the relative abundances of CS, SiO, and SiS derived by us and the very concrete physical scenario of AGB wind adopted by \cite{che2006}, it is difficult to establish meaningful conclusions regarding whether the relative abundances we derive may be ultimately driven by shock-induced chemistry in the inner wind.

It is interesting to note that regardless of which pair of molecules is chosen, the plots in Fig.~\ref{fig:ratios} show a trend in which the higher the abundance of one molecule the more abundant the other is. That is, the abundances of CS, SiO, and SiS seem to scale with each other, so that there are envelopes in which the three molecules are abundant, while in others the three molecules maintain low abundance levels. This conclusion seems robust when considering CS and SiO, although it may not be completely true concerning SiS because in some of the sources where SiS is not detected, SiS may have a low abundance while CS and SiO do not. This point is discussed in more detail below. We note that a correlation of this type was found by \cite{dan2018} for CS and SiS in a sample including C-, M-, and S-type stars, although in that study the trend is considered tentative because of the small number of sources included.

In Fig.~\ref{fig:abun} we plot the fractional abundances $f_0$ derived for CS, SiO, and SiS as a function of the density in the envelope, evaluated through the quantity $\dot{M}/V_{\rm exp}$. In the case of CS, the data strongly suggest a negative correlation between CS abundance and envelope density. The same behavior is even more evident for SiO. That is, as the density in the envelope increases the abundances of CS and SiO decrease. This behavior was already found for SiO in both M-type stars by \cite{gon2003} and carbon stars by \cite{sch2006a} and was interpreted as an evidence of enhanced SiO adsorption onto dust grains (and thus depletion from the gas phase) with increasing density. Using a larger sample of carbon stars, we thus confirm the result found by \cite{sch2006a} for SiO. The same trend of decreasing abundance with increasing density was also found recently for SiC$_2$, pointing to this molecule as a gas-phase precursor of silicon carbide dust around carbon stars \citep{mas2018}. We note that in the recent study by \cite{dan2018}, the abundances of CS derived in C-type AGB envelopes do not show the anticorrelation with envelope density found by us. The reason is that their sample of carbon stars, with just seven objects, is small and the range of $\dot{M}/V_{\rm exp}$ covered does not include high density envelopes, of which  those in our sample give better evidence of the trend of decreasing abundance with increasing density.

It is worth looking at the predictions of TE to investigate whether the anticorrelation between abundance and envelope density observed for CS and SiO could be caused by an effect of the density on the TE abundance of these molecules. As shown in Fig.~\ref{fig:lte_abun}, the main effect of an increase in the mass loss rate, and thus on the envelope density, on the abundances of CS and SiO is that the curves shift slightly to inner radii. If the radius at which the CS abundance is quenched to its TE value is the same for different mass loss rates, then the quenched abundance of CS would be lower for higher mass loss rates, in agreement with the observed behavior. It is however unlikely that the radius of abundance quenching is the same for different mass loss rates because higher densities would make the region of abundance quenching occur at larger radii. The reason is that higher densities imply shorter chemical timescales and a larger region in which TE prevails. This would result in an even more pronounced decrease of the quenched CS abundance with increasing density. Although qualitatively this scenario would be in agreement with the observed trend of decreasing abundance of CS with increasing density, a detailed chemical kinetics model is needed to obtain quantitative estimates. In the case of SiO, a similar reasoning implies that the quenched abundance would be higher for higher densities (see \citealt{agu2006}) contrary to what observations indicate. It is therefore very unlikely that the observed decrease in the abundance of SiO with increasing density is caused by gas-phase chemistry. The most likely explanation is that SiO incorporates into dust grains and depletes from the gas phase with increasing density in the envelope.

\begin{figure*}
\centering
\includegraphics[width=0.95\columnwidth]{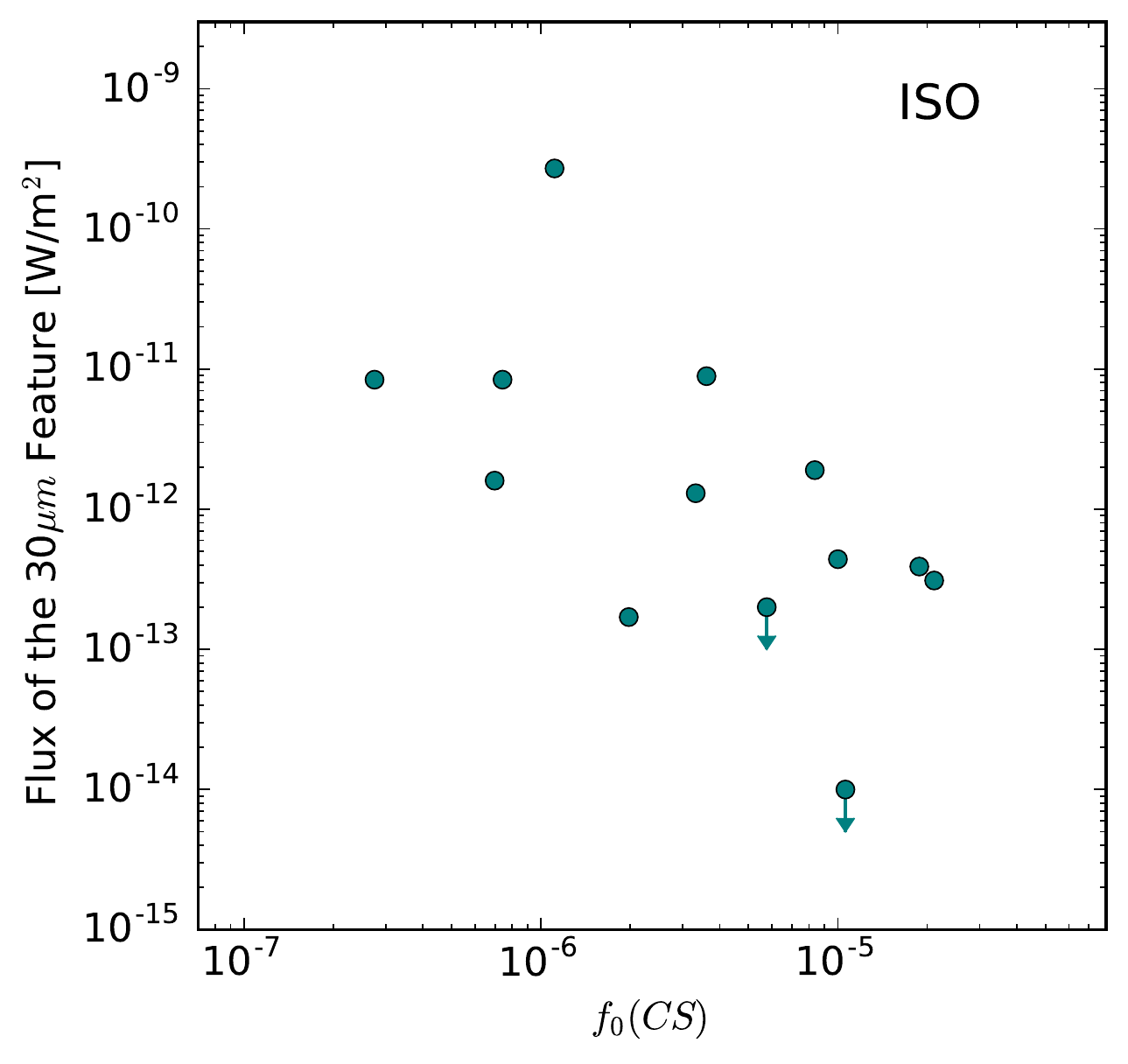}
\includegraphics[width=0.95\columnwidth]{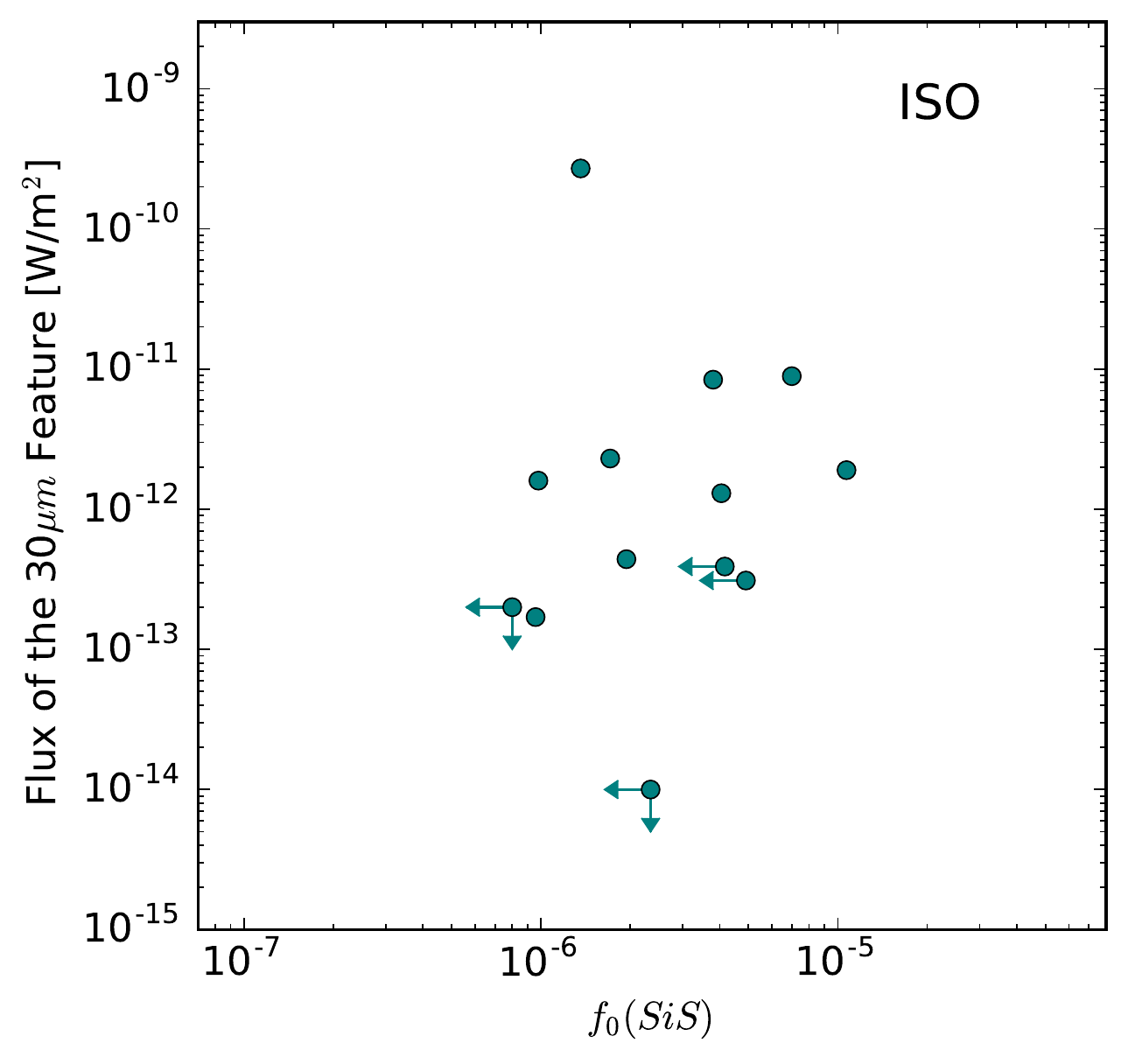}
\caption{Flux of the MgS dust feature at 30\,$\muup$m observed by ISO \citep{hon2002} vs. the fractional abundance of CS (left panel) and SiS (right panel) derived in this work.} \label{fig:mgs}
\end{figure*}

The case of SiS deserves special attention. It is remarkable that SiS is not detected in envelopes with low mass loss rates below $10^{-6}$ M$_{\odot}$ yr$^{-1}$, while it is detected in all sources with mass loss rates above that threshold. This fact, which has been noticed to some extent in previous observational studies \citep{buj1994,sch2007,dan2018}, is shown in this work in a robust way (see Table~\ref{table:abundances} and right panel of Fig.~\ref{fig:abun}). The reason behind this behavior is not clear. The non-detection of SiS in these envelopes might either be due to a lack of sensitivity or a low fractional abundance of the molecule. Looking at the predictions of TE (right panel in Fig.~\ref{fig:lte_abun}), we note that for low mass loss rates the predicted abundance of SiS is indeed lower. This  fact is even more marked if we take into account that the radius at which the abundance quenches to the TE value is expected to be shifted to inner radii for lower mass loss rates. A scenario of TE chemistry plus abundance quenching would be in agreement with objects with low mass loss rates having lower SiS abundances. If that is the underlying cause, it is however strange that the observed SiS abundances do not show a smooth variation with density but an abrupt differentiation between sources with and without SiS detection at $10^{-6}$ M$_{\odot}$ yr$^{-1}$, and show no evidence of increasing SiS abundance with increasing density for those sources where SiS is detected. We suspect that the lack of SiS detections in objects with low mass loss rates could be caused by a lack of its constituent elements, which would be largely trapped in SiO and SiC$_2$ (in the case of silicon) and CS (regarding sulfur). As shown in this work and in \cite{mas2018}, the molecules SiO, CS, and SiC$_2$ (probably also Si$_2$C; \citealt{cer2015}) become very abundant in C-rich objects with low mass loss rates. This suspicion however would need to be corroborated with a detailed chemical kinetics model of the inner wind of envelopes with different mass loss rates. Apart from the upper limits to the abundance of SiS in objects with low mass loss rates, the SiS abundances derived through positive detections show hints of decreasing SiS abundance with increasing density (right panel in Fig.~\ref{fig:abun}). This has to be seen as tentative, however, and in any case it is not as evident as in the cases of CS and SiO. Therefore, if the tentative decrease in the abundance of SiS with increasing envelope density is interpreted in terms of adsorption onto dust grains, we can conclude that SiS is not incorporated into dust grains to an extent as important as that of SiO. 

\cite{sch2007} found that the SiS abundance does not show any particular correlation with the envelope density for C-rich envelopes. However, they found a slightly better fit to their observations for the case of IRC\,+10216 by including a compact SiS component with a fractional abundance $1.7\times10^{-5}$ out to a radius of $5\times10^{14}$ cm, which could imply an SiS abundance gradient in line with the results found by \cite{agu2012}. Introducing an inner high-abundance SiS component also produced a better fit for the oxygen-rich IK\,Tau implying SiS depletion. This result is similar to that found by \cite{dec2010a} when modeling low- and high-excitation lines, however a recent study by \cite{dan2019}, using sensitive ALMA observations to determine the SiS distribution in the envelope of IK\,Tau, does not reveal a compact inner region as previously found. Regardless, we do maintain that the evidence of SiS depletion in C-rich envelopes is weak.

In any case, in this work we investigated SiS in a larger sample of carbon stars covering a broader range of envelope densities than previously studied, which permitted us to clearly see a systematic non-detection of SiS at low mass loss rates and a tentative negative correlation between SiS abundance and envelope density seen at high mass loss rates. 

\subsection{MgS dust: Possible gas-phase precursors}

A common characteristic that is seen in C-rich evolved stars is the presence of a prominent IR emission band that is centered around 30\,$\muup$m. The feature was discovered by \cite{for1981} in dusty carbon-rich environments; \cite{goe1985} first suggested that magnesium sulfide (MgS) dust may be responsible for this spectral feature and this suggestion has been widely accepted since then. However, the carrier is still argued upon. One of the major concerns is regarding the sulfur abundance that is required to explain the observed emission. \cite{zha2009} argued that the amount of MgS required to explain the power emitted by the 30\,$\muup$m feature in the post-AGB, HD\,56126, would require ten times more atomic sulfur than available in the ejected envelope. \cite{zhu2008} found the only way MgS dust can form for C-rich AGB stars is by precipitation on preexisting silicon carbide grains. \cite{lom2012} then discussed that if the grains were of a heterogeneous composite nature, meaning if MgS dust forms in a layer coated around an amorphous carbon/SiC core grain, then there would not be an abundance constraint. Investigating the formation mechanism of MgS dust is out of the scope of this paper. However, we can investigate if there is a connection between the sulfur bearing molecules studied in this work and the 30\,$\muup$m feature attributed to MgS dust. This way we could identify if any of these S-bearing molecules could act as precursors of MgS dust in the ejecta of AGB stars. More specifically, we aim to investigate whether there is a relation between the derived fractional abundances of CS and SiS and the strength of the 30\,$\muup$m feature.

\cite{hon2002} carried out an observational study of the 30\,$\muup$m feature of a large number of C-rich sources observed with the Infrared Space Observatory (ISO). There are 13 sources in common between their sample and ours. In the left panel of Fig.~\ref{fig:mgs} we plot the flux of MgS dust versus the fractional abundance of CS for these 13 sources. We clearly see a trend in which the flux of MgS dust increases as the gas-phase abundance of CS decreases. If the flux of the 30\,$\muup$m feature is a good proxy of the amount of MgS dust and if the hypothesis that CS is a gas-phase precursor of dust in C-rich AGB stars is correct, then this result supports the idea that the decline in the abundance of CS with increasing envelope density is caused by a more efficient incorporation of CS on MgS dust. In the case of SiS there is no obvious correlation between the fractional abundance and the flux of the 30\,$\muup$m feature (see right panel of Fig.~\ref{fig:mgs}). The lack of correlation suggests that SiS does likely not play a role in the formation of MgS dust in these stars, although we note that the range of SiS abundances covered is limited. \cite{smo2012} carried out an observational study on a large sample of S-type stars and found that many stars that show the MgS emission feature also show emission peaks at  6.6\,$\muup$m and 13.5\,$\muup$m due to molecular SiS. This fact led these authors to suggest that MgS dust could be formed as a consequence of a reaction between Mg and SiS in S-type stars in contrast with our conclusion that CS, rather than SiS, is a precursor of MgS dust in C-rich AGB stars.

\subsection{SiO and SiS as possible precursors of SiC dust}

\begin{figure*}
\centering
\includegraphics[width=0.95\columnwidth]{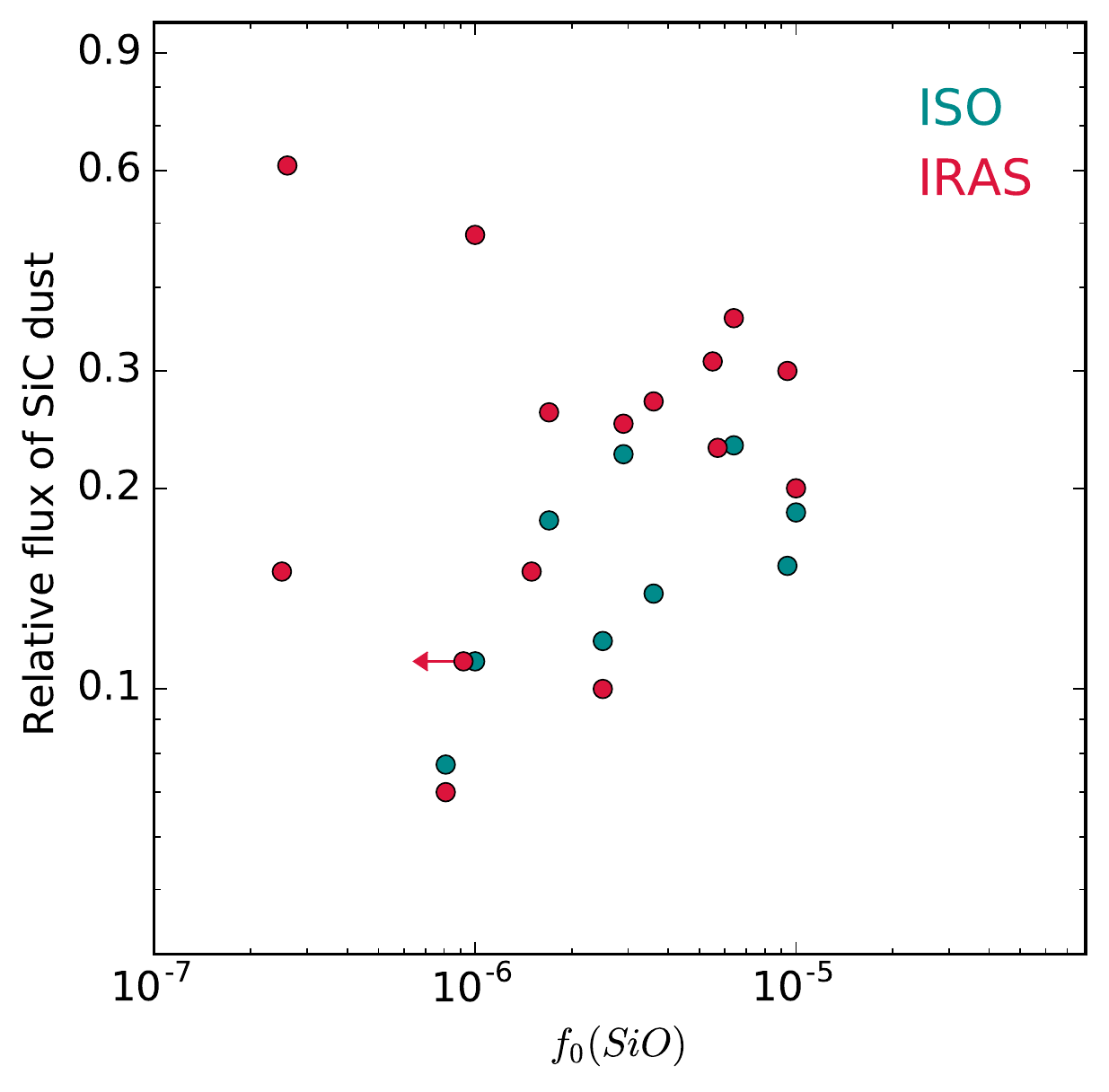}
\includegraphics[width=0.95\columnwidth]{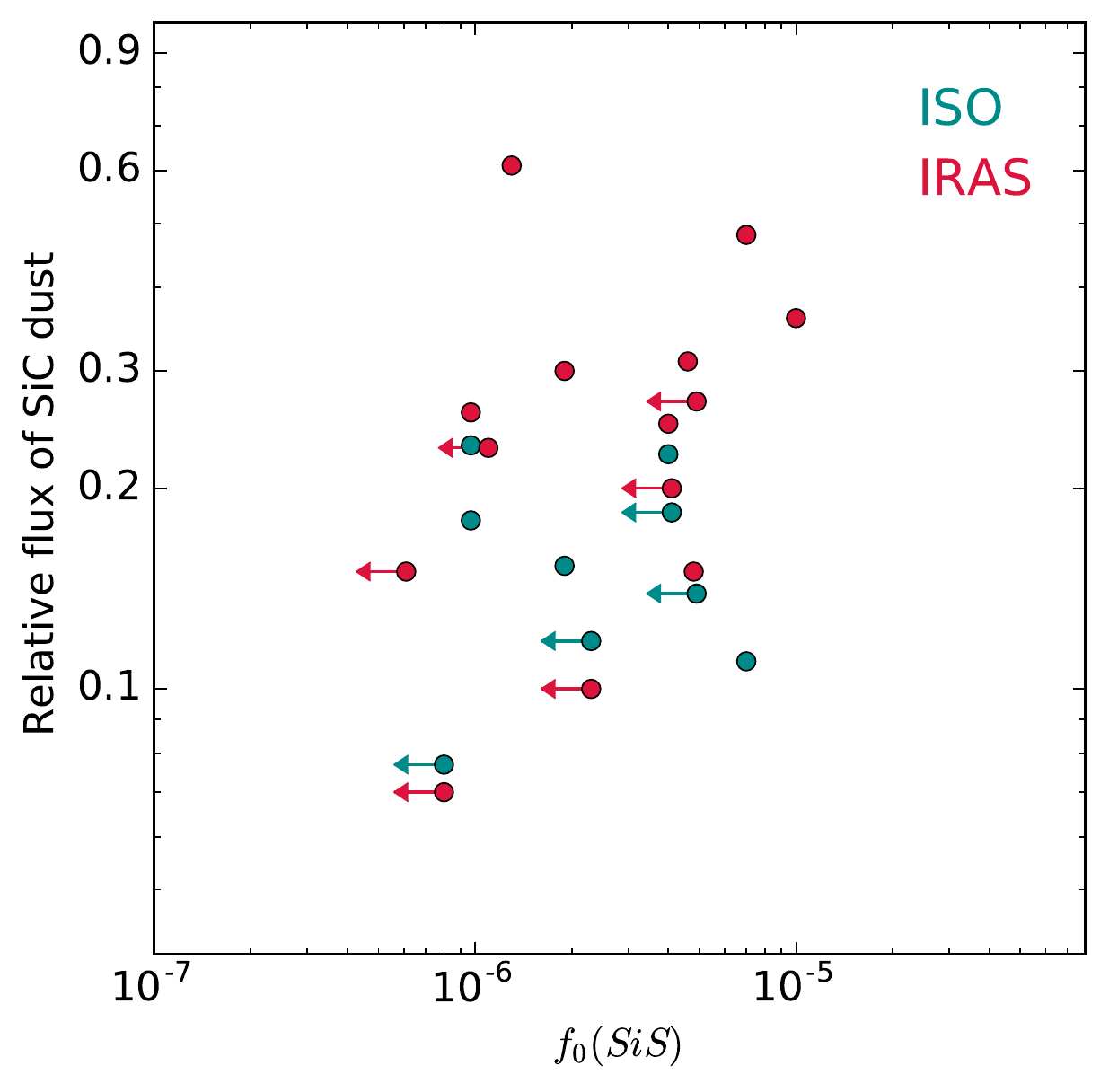}
\caption{Relative integrated flux of the SiC dust feature at 11.3\,$\muup$m taken from \cite{slo98} and \cite{yan2004} vs. the fractional abundance of SiO and SiS derived in this work.} \label{fig:sic}
\end{figure*}

Since both SiO and SiS are important Si-carriers (\citealt{olo1982,luc1995,agu2012}), in this section we assess if there could be a relation between these molecules and the formation of SiC dust around AGB stars. We collected information on the Infrared Astronomical Satellite (IRAS) and the ISO data for the sources in our sample that exhibit the SiC dust emission feature at 11.3\,$\muup$m. \citet{slo98} analyzed the IRAS spectra of carbon stars with which we share 15 sources, and \citet{yan2004} studied the ISO spectra of C-rich AGB stars with 9 sources in common with our sample. These authors determined the relative flux of SiC dust as the ratio of the integrated flux of the 11.3\,$\muup$m emission feature (after continuum substraction) divided by the integrated flux of the continuum. In Fig.~\ref{fig:sic} we plot the relative integrated flux of SiC dust versus the fractional abundance of SiO (left) and SiS (right) for the sources in our sample that have IRAS or ISO data. With the same reasoning as in the previous section, if the relative flux of SiC dust is a direct indicator of the amount of silicon carbide dust and the hypothesis that SiO and/or SiS contribute to the formation of SiC dust is correct, we would expect to see a trend in which the relative flux of SiC dust increases as the gas-phase abundance of SiO and/or SiS decreases. However, Fig.~\ref{fig:sic} does not show any clear trend indicating this; the same result was found between the gas-phase SiC$_2$ and SiC dust by \cite{mas2018}. Regardless, it is important to note that the relative flux of the 11.3\,$\muup$m SiC band is an observable quantity that may not be a direct indicator of the mass of silicon carbide dust in the envelope. The derivation of the amount of SiC dust in the envelope requires a radiative transfer analysis that involves a meticulous description of the chemical composition, size, and temperature of dust throughout the envelope. 

\subsection{Radial extent: Photodissociation versus empirical relations} \label{sec:photodissociation}

As discussed in Sec.~\ref{subsec:emission_size}, the radial extent of CS, SiO, and SiS in carbon star envelopes is probably controlled by photodissociation from the ambient interstellar UV radiation field. However, given the aforementioned difficulties to account properly for the abundance falloff using photodissociation rates from the literature (in which case the radial extent is very likely underestimated), in this work we adopted a simple radial abundance distribution given by a Gaussian function with $e$-folding radii derived from empirical relations between $r_e$ and the wind density. These relations were ultimately based on a scaling law between $r_e$ and $\dot{M}$/$V_{\rm exp}$ derived from a multiline study of SiO in M-type stars \citep{gon2003}. In this work we adopted the same relation for SiO, based on the assumption that SiO behaves similarly in envelopes around M- and C-type stars; for SiS we adopted the same relation, based on the assumption that SiS and SiO have similar emission sizes as mentioned previously; and for CS we adopted a modified version, based on arguments to avoid requiring a sulfur abundance higher than the solar abundance. If we consider as good the adopted empirical relations and we assume that the abundance falloff is entirely controlled by photodissociation, then it is possible to extract some useful conclusions about the photodissociation of these molecules. More specifically, we aim to find the unattenuated photodissociation rates that best reproduce the abundance falloff given by the adopted empirical laws for $r_e$.

\begin{table}
\caption{Photodissociation parameters}\label{table:rates}
\centering
\begin{tabular}{lccc|cc}
\hline \hline
Molecule & $\alpha$ & $\beta$ & Ref. & $\alpha'$ & $\langle$Dev.$\rangle$ \\
\hline
\rule{0pt}{2.5ex}
CS & 3.7 $\times$ 10$^{-10}$ & 2.32 & a & $1.5 \times 10^{-10}$ & 35\% \\
\hline
\rule{0pt}{2.5ex}
SiO & 1.6 $\times$ 10$^{-9}$ & 2.66 & b & $7 \times 10^{-10}$ & 31\% \\
\hline
\rule{0pt}{2.5ex}  
SiS & 1.6 $\times$ 10$^{-9}$ & 2.66 & c & $8 \times 10^{-10}$ &  47\% \\
\hline
\end{tabular}
\tablenotea{The values of $\alpha$ and $\beta$ are taken from the literature. References are (a) \cite{pat2018}, (b) \cite{hea2017}, (c) assumed the same as SiO. The  parameter $\alpha'$ are the values that best reproduce the empirical relations between $r_e$ and $\dot{M}$/$V_{\rm exp}$ adopted in this study. The mean deviations $\langle$Dev.$\rangle$ are also given (see text).}
\end{table}

\begin{figure}
\centering
\includegraphics[width=0.85\columnwidth]{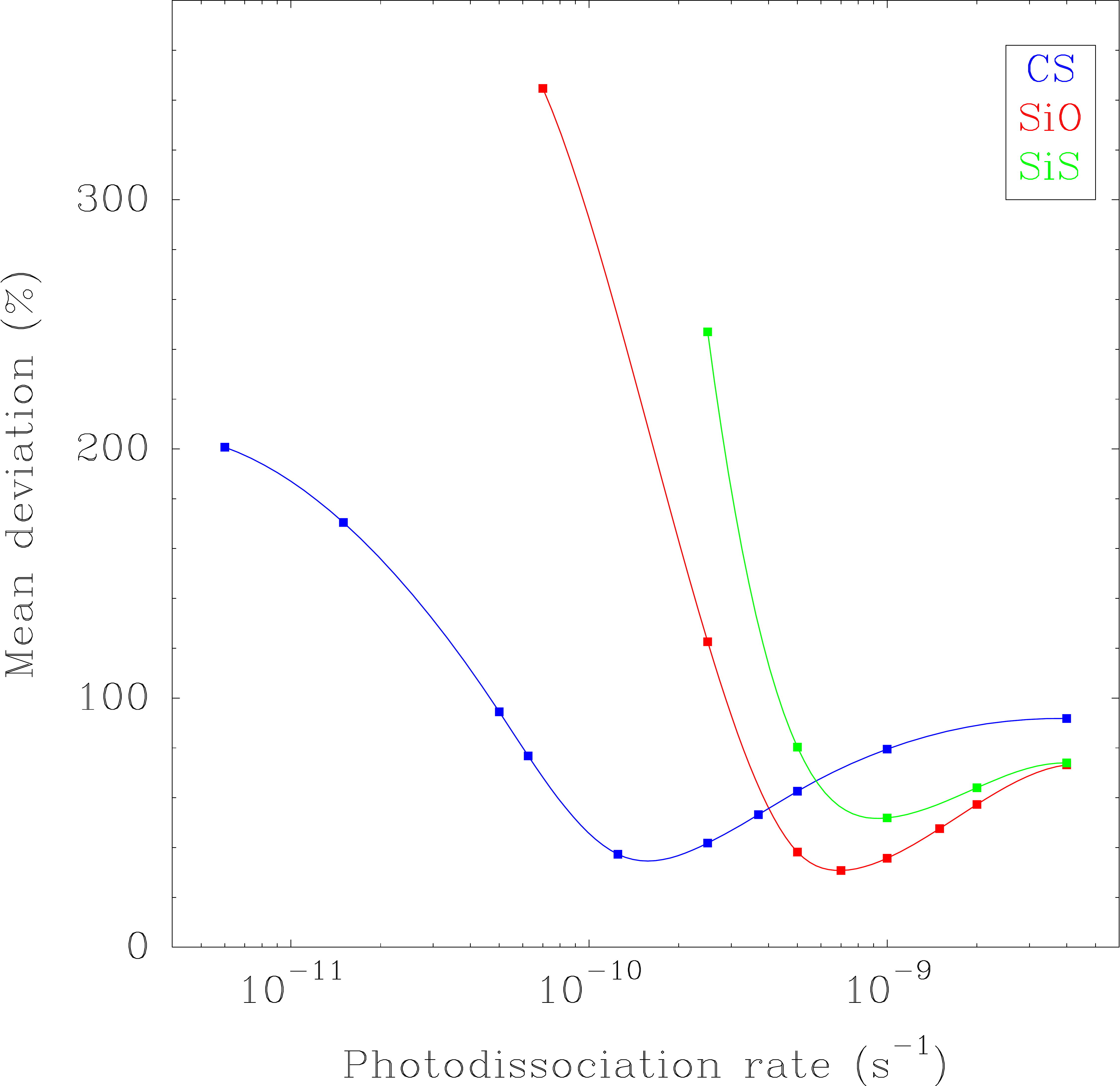}
\caption{The mean deviation between the calculated line areas with the photodissociation model and the model using empirical laws is plotted as a function of the unattenuated photodissociation rate adopted in the photodissociation model for CS, SiO, and SiS.}
\label{fig:photodissociation_rate}
\end{figure}

The photodissociation model used for this exercise is the same employed in \cite{mas2018}. Briefly, the radial variation of the fractional abundance can be expressed as \citep{jur1981,hug1982}
\begin{equation}
\frac{df}{dr} = - \frac{\alpha}{V_{\rm exp}} \exp \left [-\left (\frac{r_d}{r} \right) \right] f,
\end{equation} 
where $\alpha$ is the unattenuated photodissociation rate in s$^{-1}$ and the photodissociation radius $r_d$ is given by
\begin{equation}
r_d = \frac{\beta\, \dot{M}}{4 \pi\, V_{\rm exp}\, \overline{m}_{g}\, 1.87\times 10^{21}},
\end{equation} 
where $\beta$ is the dust shielding factor, $\overline{m}_{g}$ is the average mass of gas particles in grams, and the numerical value is the canonical $N_{\rm H}$/$A_V$ ratio for the local ISM \citep{boh1978}. In Table~\ref{table:rates} we list the unattenuated photodissociation rates $\alpha$ and dust shielding factors $\beta$ collected from the literature for CS, SiO, and SiS, which were used as starting values. For each of the three molecules, we then varied the unattenuated photodissociation rate in small steps, computed the associated abundance radial profile for each envelope in the sample, and calculated the corresponding line profiles. To decide which photodissociation rate agrees best with the empirical abundance falloff, we computed for each envelope the difference between the velocity-integrated line intensity calculated with the photodissociation model and that calculated with the empirical falloff model and then estimated the mean difference.

The mean deviations found are plotted as a function of the photodissociation rate in Fig.~\ref{fig:photodissociation_rate}. The optimal photodissociation rate $\alpha'$ for which the lowest mean deviation is found is given in the right side of Table~\ref{table:rates} for each of the three molecules, together with the associated mean deviation. In the case of CS, we found that the lowest mean deviation (35 \%) occurs for $\alpha'=1.5\times10^{-10}$ s$^{-1}$, while for SiO and SiS the lowest deviation (31 and 47 \%, respectively) is found for photodissociation rates of $7\times10^{-10}$ and $8\times10^{-10}$ s$^{-1}$, respectively. There are two interesting aspects to comment on. First, the optimal photodissociation rates found for SiO and SiS are very similar, while that of CS is significantly lower. These results merely reflect the adopted empirical abundance falloffs, which were the same for SiO and SiS, while for CS we adopted an empirical scaling law implying an outer abundance falloff. And second, the optimal photodissociation rates $\alpha'$ are significantly lower than the literature values $\alpha$. This illustrates in a different way our initial suspicion that literature photodissociation rates underestimate the radial extent of CS, SiO, and SiS. Whether, this finding implies that real photodissociation rates are indeed lower than the literature values or whether this points to a different $N_{\rm H}$/$A_V$ ratio than the canonical interstellar value of \cite{boh1978} is not yet clear.

\section{Conclusions} \label{sec:conclusions}

In this work we used the IRAM 30 m telescope to survey a sample of 25 C-rich CSEs in the $J=3-2$ line of CS and SiO and in the $J=7-6$ and $J=8-7$ lines of SiS. We carried out excitation and radiative transfer calculations based on the LVG method to derive fractional abundances relative to H$_2$. We found that the abundances of the three molecules are positively correlated with each other (especially those of CS and SiO) and that while CS and SiS have similar abundances, SiO is present at a lower abundance level.

We also found a clear trend in which SiO and CS become less abundant as the envelope density increases. The depletion of CS with increasing density can be due to gas-phase chemistry in the inner wind or to incorporation onto dust grains. The latter scenario is favored by the fact that we find a negative correlation between the fractional abundance of CS and the 30\,$\muup$m feature attributed to MgS dust, which suggests that CS is a likely precursor of MgS dust grains in C-rich AGB envelopes. In the case of SiO, the most likely explanation of the negative correlation between fractional abundance and envelope density is that SiO incorporates more efficiently onto dust grains at increasing density owing to the enhanced collision rate between particles and the acceleration of accretion and coagulation processes. Thus, both CS and SiO are probable candidates to act as gas-phase precursor of dust grains.

We find that SiS is systematically not detected in envelopes with mass loss rates below $10^{-6}$ M$_{\odot}$ yr$^{-1}$, probably because of the increasing importance of other molecules that lock most silicon and sulfur (SiO, SiC$_2$, and CS) or because of a lack of sensitivity. The SiS abundances derived in the sources in which the molecule is detected suggest a tentative trend of decreasing abundance with increasing density.  This  trend indicates that SiS could also be incorporated into dust grains, although the non-detections clearly undermine this tentative trend. Nevertheless, this conclusion is not as robust as that of CS and SiO.

Our conclusions on the role of CS, SiO, and SiS as gas-phase precursors of dust are based on low energy lines, which probe post-condensation regions. More observations, in particular high-J lines and interferometric observations probing the inner regions of the envelopes, are needed to affirm the conclusions obtained in this study.

\begin{acknowledgements}

We are grateful to Otoniel Denis-Alpizar, Christian Balan{\c c}a, and Fabrice Dayou for providing the latest collisional rate coefficients of CS and SiO. We thank the IRAM 30 m staff for their help during the observations. This research has made use of the SIMBAD database, operated at CDS, Strasbourg, France. We acknowledge funding support from the European Research Council (ERC Grant 610256: NANOCOSMOS) and from Spanish MINECO through grants AYA2012-32032 and AYA2016-75066-C2-1-P. M.A. thanks Spanish MINECO for funding support through the Ram\'on y Cajal programme (RyC-2014-16277).

\end{acknowledgements}

\bibliographystyle{aa} 
\bibliography{mybib} 

\setcounter{table}{2}
\longtab{
\begin{longtable}{lcccc}
\caption{Observed line parameters of CS $J=3-2$, SiO $J=3-2$, and the $J=7-6$ and $J=8-7$ lines of SiS.}\label{table:line_parameters} \\
\hline \hline
\multicolumn{1}{l}{Line} & \multicolumn{1}{c}{$\nu_{calc}$} & \multicolumn{1}{c}{$\nu_{obs}$} & \multicolumn{1}{c}{$V_{\rm exp}$}      & \multicolumn{1}{c}{$\int T_{\rm{mb}} dv$} \\
& \multicolumn{1}{c}{(MHz)}           & \multicolumn{1}{c}{(MHz)}          & \multicolumn{1}{c}{(km s$^{-1}$)} & \multicolumn{1}{c}{(K km s$^{-1}$)} \\
\hline
\endfirsthead
\caption{Continued.} \\
\hline
\multicolumn{1}{l}{Line} & \multicolumn{1}{c}{$\nu_{calc}$} & \multicolumn{1}{c}{$\nu_{obs}$} & \multicolumn{1}{c}{$V_{\rm exp}$}  & \multicolumn{1}{c}{$\int T_{\rm{mb}} dv$} \\
& \multicolumn{1}{c}{(MHz)}  & \multicolumn{1}{c}{(MHz)}  & \multicolumn{1}{c}{(km s$^{-1}$)} & \multicolumn{1}{c}{(K km s$^{-1}$)} \\
\hline
\endhead
\hline
\endfoot
\multicolumn{5}{c}{IRC\,+10216} \\
\hline
 SiO $J=3-2$         & 130268.665  & 130268.4(1)     & 13.1(1)     &  130.7(13) \\
 SiS $J=7-6$         & 127076.180  & 127076.0(1)     & 12.7(1)     &  162.9(16)\\
 SiS $J=8-7$         & 145227.054  & 145227.0(1)     & 14.7(1)     &  270.5(27)  \\
 CS  $J=3-2$         & 146969.025  & 146968.7(1)     & 14.7(1)     &  379.3(37)\\
\hline
\multicolumn{5}{c}{CIT\,6} \\
\hline
 SiO $J=3-2$         & 130268.665  & 130267.8(5)     & 16.2(5)     & 19.8(19) \\
 SiS $J=7-6$         & 127076.180  & 127075.3(5)     & 16.4(5)     & 7.55(7)  \\
 SiS $J=8-7$         & 145227.054  & 145225.4(10)    & 17.5(10)    & 12.5(19)\\
 CS  $J=3-2$         & 146969.025  & 146968.6(5)     & 16.4(5)     & 89.0(89)\\
\hline
\multicolumn{5}{c}{CRL\,3068} \\
\hline
 SiO $J=3-2$         & 130268.665  & 130268.4(1)     & 12.3(1)    & 2.18(2)   \\
 SiS $J=7-6$         & 127076.180  & 127075.9(1)     & 12.2(1)    & 6.36(5)  \\  
 SiS $J=8-7$         & 145227.054  & 145226.7(1)     & 13.8(1)    & 9.24(9)  \\
 CS  $J=3-2$         & 146969.025  & 146968.6(1)     & 14.2(1)    & 17.2(10) \\ 
\hline
\multicolumn{5}{c}{S\,Cep} \\
\hline
 SiO $J=3-2$         & 130268.665  & 130268.6(1)    &  23.3(1)   & 7.80(7)  \\
 SiS $J=7-6$         & 127076.180  & 127076.5(10)   &  21.0(5)   & 0.30(6) $^a$ \\  
 SiS $J=8-7$         & 145227.054  & 145227.2(10)   &  20.8(10)  & 0.34(7) $^a$ \\
 CS  $J=3-2$         & 146969.025  & 146968.9(1)    &  23.8(1)   & 19.8(2) \\
\hline 
\multicolumn{5}{c}{IRC\,+30374} \\ 
\hline 
 SiO $J=3-2$         & 130268.665  & 130268.7(1)     &  25.1(2)   & 7.07(7)  \\ 
 SiS $J=7-6$         & 127076.180  & 127076.3(1)     &  22.6(4)   & 1.73(2) \\ 
 SiS $J=8-7$         & 145227.054  & 145227.2(1)     &  25.2(2)   & 2.99(3) \\
 CS  $J=3-2$         & 146969.025  & 146968.6(1)     &  25.8(2)   & 26.7(26) \\
\hline 
\multicolumn{5}{c}{LP\,And} \\ 
\hline 
 SiO $J=3-2$         & 130268.665  &  130268.5(1)    &  12.9(2)    & 8.56(8) \\ 
 SiS $J=7-6$         & 127076.180  &  127076.1(1)    &  12.1(2)    & 7.01(7)\\ 
 SiS $J=8-7$         & 145227.054  &  145226.9(1)    &  13.8(1)    & 12.3(1) \\
 CS  $J=3-2$         & 146969.025  &  146969.0(1)    &  14.7(1)    & 35.6(35) \\
\hline 
\multicolumn{5}{c}{V\,Cyg} \\ 
\hline 
 SiO $J=3-2$         & 130268.665  &  130268.5(1)    &  12.2(1)   & 9.03(9)  \\ 
 SiS $J=7-6$         & 127076.180  &  127076.4(1)    &  12.9(1)   & 1.41(1) \\ 
 SiS $J=8-7$         & 145227.054  &  145227.2(1)    &  12.3(1)   & 2.45(2) \\
 CS  $J=3-2$         & 146969.025  &  146968.8(1)    &  12.1(1)   & 25.2(25) \\
\hline
\multicolumn{5}{c}{V384\,Per} \\ 
\hline 
 SiO $J=3-2$         & 130268.665  &  130268.6(1)   & 14.8(1)    & 9.29(9) \\ 
 SiS $J=7-6$         & 127076.180  &  127076.1(1)   & 12.7(1)    & 2.15(2) \\ 
 SiS $J=8-7$         & 145227.054  &  145227.0(1)   & 14.1(1)    & 4.02(4)  \\
 CS  $J=3-2$         & 146969.025  &  146968.8(1)   & 15.6(1)    & 28.7(29) \\
\hline
\multicolumn{5}{c}{IRC\,+60144} \\ 
\hline 
 SiO $J=3-2$         & 130268.665  &  130268.6(1)   & 20.8(1)  & 5.34(5)  \\ 
 SiS $J=7-6$         & 127076.180  &  127076.2(10)  & 19.4(10) & 0.9(2) $^a$\\ 
 SiS $J=8-7$         & 145227.054  &  145226.9(1)   & 20.9(1)  & 1.65(1) \\
 CS  $J=3-2$         & 146969.025  &  146968.7(1)   & 20.9(1)  & 12.5(12) \\
\hline
\multicolumn{5}{c}{U\,Cam} \\ 
\hline 
 SiO $J=3-2$         & 130268.665  &  130268.2(1)  &  11.6(1)  & 0.96(1)  \\ 
 SiS $J=7-6$         & 127076.180  &  -  & -  & - \\ 
 SiS $J=8-7$         & 145227.054  &  -   & -  & - \\
 CS  $J=3-2$         & 146969.025  &  146968.8(1)    & 13.5(1)   & 4.7(5) \\
\hline
\multicolumn{5}{c}{IRC\,+20370} \\ 
\hline 
 SiO $J=3-2$         & 130268.665  &  130268.5(1)  & 13.1(1)  & 7.25(7)  \\ 
 SiS $J=7-6$         & 127076.180  &  127076.1(1)  & 12.7(1)  & 3.10(2) \\ 
 SiS $J=8-7$         & 145227.054  &  145227.1(5)  & 13.8(4)  & 5.64(5) \\
 CS  $J=3-2$         & 146969.025  &  146968.9(5)  & 13.2(8)  & 19.9(20) \\
\hline
\multicolumn{5}{c}{CRL\,67} \\ 
\hline 
 SiO $J=3-2$         & 130268.665  & 130268.6(1)  & 14.5(1)  & 2.88(3)  \\ 
 SiS $J=7-6$         & 127076.180  & 127076.4(1)  & 13.6(1)  & 1.62(1) \\ 
 SiS $J=8-7$         & 145227.054  & 145226.9(1)  & 15.4(2)  & 2.83(2)  \\
 CS  $J=3-2$         & 146969.025  & 146968.9(1)  & 16.1(2)  & 12.3(12) \\
\hline
\multicolumn{5}{c}{CRL\,190} \\ 
\hline 
 SiO $J=3-2$         & 130268.665  & 130269.5(5)  & 16.3(5)  & 0.39(4)  \\ 
 SiS $J=7-6$         & 127076.180  & 127076.2(1)  & 16.6(1)  & 1.45(1) \\ 
 SiS $J=8-7$         & 145227.054  & 145227.0(1)  & 16.2(1)  & 1.93(2)  \\
 CS  $J=3-2$         & 146969.025  & 146968.7(1)  & 16.9(2)  & 6.76(7) \\
\hline
\multicolumn{5}{c}{V\,Aql} \\ 
\hline 
 SiO $J=3-2$         & 130268.665  & 130268.4(1)  & 7.2(1)  & 0.44(4)  \\ 
 SiS $J=7-6$         & 127076.180  &  - & -  & - \\ 
 SiS $J=8-7$         & 145227.054  &  - & -  & -  \\
 CS  $J=3-2$         & 146969.025  & 146968.8(1)  & 9.2(1)  & 4.30(4) \\
\hline
\multicolumn{5}{c}{CRL\,2477} \\ 
\hline 
 SiO $J=3-2$         & 130268.665  & 130268.1(1)  & 16.7(1)  & 1.28(1) \\ 
 SiS $J=7-6$         & 127076.180  & 127076.2(1)  & 16.5(1)  & 2.64(2)  \\ 
 SiS $J=8-7$         & 145227.054  & 145226.8(1)  & 19.2(2)  & 2.93(3)  \\
 CS  $J=3-2$         & 146969.025  & 146968.6(1)  & 20.1(1)  & 7.17(7) \\
\hline
\multicolumn{5}{c}{CRL\,2494} \\ 
\hline 
 SiO $J=3-2$         & 130268.665  & 130268.8(1)  & 16.4(2)  & 2.64(3)  \\ 
 SiS $J=7-6$         & 127076.180  & 127077.0(10) & 20.4(10) & 1.04(20) \\ 
 SiS $J=8-7$         & 145227.054  & 145227.4(2)  & 19.0(6)  & 1.57(15)  \\
 CS  $J=3-2$         & 146969.025  & 146968.9(1)  & 19.6(1)  & 13.2(13)  \\
\hline
\multicolumn{5}{c}{Rv\,Aqr} \\ 
\hline 
 SiO $J=3-2$         & 130268.665  & 130268.3(1)  & 14.0(2)  & 5.58(5) \\ 
 SiS $J=7-6$         & 127076.180  & 127076.1(5)  & 13.3(5)  & 0.71(7) \\ 
 SiS $J=8-7$         & 145227.054  & 145226.9(1)  & 14.6(2)  & 1.34(1)  \\
 CS  $J=3-2$         & 146969.025  & 146968.6(1)  & 15.4(2)  & 13.3(13) \\
\hline
\multicolumn{5}{c}{CRL\,2513} \\ 
\hline 
 SiO $J=3-2$         & 130268.665  & 130268.5(1)  & 25.7(1)  & 2.58(2)   \\ 
 SiS $J=7-6$         & 127076.180  & 127076.0(5)  & 24.2(6)  & 0.84(8) \\ 
 SiS $J=8-7$         & 145227.054  & 145226.7(5)  & 24.9(4)  & 1.66(16)  \\
 CS  $J=3-2$         & 146969.025  & 146968.7(1)  & 26.5(2)  & 8.92(9) \\
\hline
\multicolumn{5}{c}{S\,Aur} \\ 
\hline 
 SiO $J=3-2$         & 130268.665  & 130267.5(1)  & 21.8(2)  & 1.25(1)  \\ 
 SiS $J=7-6$         & 127076.180  &  - & -  & - \\ 
 SiS $J=8-7$         & 145227.054  & -  & -  & -  \\
 CS  $J=3-2$         & 146969.025  & 146968.1(1)  & 26.5(1)  & 4.05(4) \\
\hline
\multicolumn{5}{c}{V636\,Mon} \\ 
\hline 
 SiO $J=3-2$         & 130268.665  & 130269.3(1)   & 24.3(1)  & 3.74(4)  \\ 
 SiS $J=7-6$         & 127076.180  & 127076.2(5)   & 24.1(5)  & 0.60 $^a$ \\ 
 SiS $J=8-7$         & 145227.054  & 145227.9(5)  & 26.5(5)  &  0.50(5) $^a$ \\\
 CS  $J=3-2$         & 146969.025  &  146969.8(1)  & 25.8(1)  & 8.85(9) \\
\hline
\multicolumn{5}{c}{W\,Ori} \\ 
\hline 
 SiO $J=3-2$         & 130268.665  & 130268.1(5)   & 8.5(4)  & 0.29(3)  \\ 
 SiS $J=7-6$         & 127076.180  & - & - & - \\ 
 SiS $J=8-7$         & 145227.054  & -  & -  & - \\\
 CS  $J=3-2$         & 146969.025  & 146968.8(5)  & 10.5(8)  & 4.2(4) \\
\hline
\multicolumn{5}{c}{Y\,CVn} \\ 
\hline 
 SiO $J=3-2$         & 130268.665  & 130268.3(1)   & 7.3(1)   & 0.38(4) \\ 
 SiS $J=7-6$         & 127076.180  & - & - & - \\ 
 SiS $J=8-7$         & 145227.054  & -  & -  & - \\\
 CS  $J=3-2$         & 146969.025  & 146968.5(1)  & 9.4(1)  & 7.58(7) \\
\hline
\multicolumn{5}{c}{R\,Lep} \\ 
\hline 
 SiO $J=3-2$         & 130268.665  & 130267.9(1)  & 19.8(2)  & 3.45(3)  \\ 
 SiS $J=7-6$         & 127076.180  & - & - & -\\ 
 SiS $J=8-7$         & 145227.054  & - & -  &   \\
 CS  $J=3-2$         & 146969.025  & 146968.3(1)  & 20.9(2)  & 5.78(6)  \\
\hline
\multicolumn{5}{c}{ST\,Cam} \\ 
\hline 
SiO $J=3-2$          & 130268.665  & -  & - & - \\ 
 SiS $J=7-6$         & 127076.180  & -  & -  & - \\ 
 SiS $J=8-7$         & 145227.054  & - & -  & -  \\
 CS  $J=3-2$         & 146969.025  & 146969.2(1)  & 11.5(1)  & 0.82(8)  \\
\hline 
\multicolumn{5}{c}{UU\,Aur} \\ 
\hline 
 SiO $J=3-2$         & 130268.665  & 130266.2(10)  & 6.7(10)  & 0.09(2)  \\ 
 SiS $J=7-6$         & 127076.180  & -  & -  & - \\ 
 SiS $J=8-7$         & 145227.054  & -  & -  & -  \\
 CS  $J=3-2$         & 146969.025  &  146968.6(10) & 11.0(1)  & 0.30(3)  \\
\hline
\end{longtable}
\tablefoot{
Numbers in parentheses are 1$\sigma$ uncertainties in units of the last digits.\\
$^a$ Marginal detection. \\
}}

\end{document}